\def\Im{{\text{Im}}\,}
\def\Re{{\text{Re}}\,}

\def\be{\begin{equation}}
\def\ee{\end{equation}}
\def\bea{\begin{eqnarray}}
\def\eea{\end{eqnarray}}
\def\bse{\begin{subequations}}
\def\ese{\end{subequations}}

\documentclass[prb,twocolumn,showpacs,amsmath,amssymb,eqsecnum,preprintnumbers]{revtex4}
\usepackage{graphicx}
\usepackage{dcolumn}
\usepackage{bm}
\begin{document}
\title{Theory of helimagnons in itinerant quantum systems II: \\
       Nonanalytic corrections to Fermi-liquid behavior}
\author{D. Belitz$^{1,2}$, T. R. Kirkpatrick$^{1,3}$, and A. Rosch$^4$}
\affiliation{$^{1}$ Kavli Institute for Theoretical Physics, University of
                    California, Santa Barbara, CA 93106, USA\\
             $^{2}$ Department of Physics and Materials Science Institute, University
                    of Oregon, Eugene, OR 97403, USA\\
             $^{3}$ Institute for Physical Science and Technology,
                    and Department of Physics, University of Maryland, College Park,
                    MD 20742, USA\\
             $^{4}$ Institut f{\"u}r Theoretische Physik, Universit{\"a}t zu K{\"o}ln,
                    Z{\"u}lpicher Strasse 77, D-50937 K{\"o}ln, Germany}
\date{\today}

\begin{abstract}
A recent theory for the ordered phase of helical or chiral magnets such as MnSi
is used to calculate observable consequences of the helical Goldstone modes or
helimagnons. In systems with no quenched disorder, the helimagnon contributions
to the specific heat coefficient is shown to have a linear temperature
dependence, while the quasi-particle inelastic scattering rate is anisotropic
in momentum space and depends on the electronic dispersion relation. For cubic
lattices the generic temperature dependence is given by a non-Fermi-liquid
$T^{3/2}$ behavior. The contribution to the temperature dependence of the
resistivity is shown to be $T^{5/2}$ in a Boltzmann approximation. The
helimagnon thus leads to nonanalytic corrections to Fermi-liquid behavior.
Implications for experiments, and for transport theories beyond the Boltzmann
level, are discussed.
\end{abstract}


\maketitle

\section{Introduction}
\label{sec:I}

The itinerant helical magnet MnSi has attracted a considerable amount of
interest lately. This material shows, at ambient pressure, helical magnetic
order below a critical temperature $T_{c}\approx 30\,{\text
K}$.\cite{Ishikawa_et_al_1976} The wavelength of the helix is $2\pi /q$, with
$q\approx 0.035\,\text{\AA}$ the pitch wave number.\cite{Ishikawa_et_al_1976}
Application of hydrostatic pressure $p$ monotonically decreases $T_{c}$ until
$T_{c}$ vanishes at $p = p_{c}\approx 14\,
\text{kbar}$.\cite{Pfleiderer_et_al_1997} A tricritical point is observed on
the phase boundary at $T\approx 10\,\text{K}$, such that the
paramagnetic-to-helimagnetic transition at higher pressures and lower
temperatures is of first order, while the transition at lower pressures and
higher temperatures is of second or very weakly first
order.\cite{Pfleiderer_et_al_1997, Pfleiderer_Julian_Lonzarich_2001} In the
ordered phase, neutron scattering shows the helical pattern of the
magnetization with the axis of the helix pinned in the $[111]$-direction due to
crystal field effects.\cite{Ishikawa_Arai_1984} In the disordered phase,
quasi-static remnants of helical order are still observed at low temperatures
close to the phase boundary. The distribution of the helical axis orientation
is much more isotropic than in the ordered phase, with broad maxima in the
$[110]$-direction.\cite{Pfleiderer_et_al_2004} Such remnants of order in the
disordered phase are not entirely unexpected close to a first order phase
transition boundary.\cite{Chaikin_Lubensky_1995} In the entire disordered
phase, up to a temperature of a few Kelvin, and up to the highest pressures
investigated ($\approx 2p_{c}$), pronounced non-Fermi-liquid behavior of the
resistivity is observed, with the temperature dependence of the resistivity
given by $\rho (T\rightarrow 0)\sim {\text{const.}} +
T^{3/2}$.\cite{Pfleiderer_Julian_Lonzarich_2001} In the ordered phase, on the
other hand, the transport is observed to be Fermi-liquid-like, with a leading
$T^{2}$-dependence of the resistivity. No explanation has been given so far for
the unusual properties in the disordered phase. However, it is natural to
speculate that the remnants of the helical order that are clearly observed in
the paramagnetic phase have something to do with them. As a first step in
investigating this possibility, the effects of the helical magnetic order on
the electronic properties in the ordered phase of an itinerant electron system
need to be understood.

In a previous paper, hereafter denoted by I, we presented a theory for the
long-ranged order and fluctuations in the helically ordered phase of itinerant
chiral magnets.\cite{paper_I} In particular, we obtained the helical Goldstone
modes or helimagnons in the ordered phase. Physical quantities computed
included the leading contribution to the dynamical magnetic susceptibility at
wave numbers near the helical pitch wave number, and the noninteracting
single-particle Green function in the ordered phase. Results for the helimagnon
consistent with I were obtained in Ref.\ \onlinecite{Maleyev_2005}.

In the present paper we use the results of I to determine some of the
thermodynamic and transport properties of the helical magnetic state. In
particular we will calculate the effects of the magnetic fluctuations in the
ordered phase on the specific heat coefficient and on the electrical
resistivity to first order in these fluctuations. We will see that the magnetic
soft modes lead to nonanalytic corrections to the standard Fermi-liquid theory
results. Specifically, we find that the specific heat coefficient $\gamma$ has
a contribution linear in the temperature $T$, whereas Fermi-liquid theory gives
a leading correction to the constant Pauli value that is proportional to
$T^{2}\ln T$. Similarly, the resistivity we find to have a $T^{5/2}$ term while
Fermi-liquid theory gives a $T^{3}$ correction to the leading $T^{2}$
contribution. Finally, the quasi-particle or inelastic relaxation rate has a
temperature dependence proportional to $T^{3/2}$, which is stronger than the
$T^2$ behavior found in Fermi-liquid theory. Similar effects in metallic
antiferromagnets, in particular an anomalously large scattering rate, have been
discussed in Ref.\ \onlinecite{Irkhin_Katsnelson_2000}

The plan of this paper is as follows. In Section II we give simple physical
plausibility arguments that show how our results arise and how they are related
to the behavior of more familiar systems. In Section III the results are
derived using a combination of field theory, many-body perturbation theory, and
transport theory. In Section IV we further discuss our results and their
experimental implications.

\section{Simple physical arguments, and results}
\label{sec:II}

In this Section we use simple physical arguments to obtain some of our results
and discuss certain aspects of them. More detailed and thorough technical
derivations will be given in later sections.

\subsection{Specific heat}
\label{subsec:II.A}

In I we showed that in the chiral magnetic state there is one Goldstone mode,
the {\em helimagnon}. Denoting the frequency of the helimagnon by $\omega_0$
and the wave vector by ${\bm k}$, the dispersion relation in the
long-wavelength limit is given by\cite{rot_invariance_footnote}
\be
\omega_0({\bm k}) = \sqrt{c_{z}k_z^2 + c_{\perp}{\bm k}_{\perp}^4},
\label{eq:2.1}
\ee
where $c_{z}$ and $c_{\perp}$ are elastic constants which can be expressed in
terms of the exchange splitting, the pitch wave number, and the Fermi wave
number. The wave vector ${\bm k} = ({\bm k}_{\perp},k_{z})$ has been separated
into longitudinal and transverse components with respect to the pitch wave
vector ${\bm q} = (0,0,q)$, which we take to point in the $z$-direction. Notice
that this dispersion relation is strongly anisotropic, and softer in the
direction transverse to the pitch vector, $\omega_0 \sim {\bm k}_{\perp}^2$ for
$k_z = 0$, than in the $z$-direction, $\omega_0 \sim k_z$ for $k_{\perp} \equiv
\vert{\bm k}_{\perp}\vert = 0$.

To the extent that the helimagnon is a well-defined quasi-particle, one expects
its contribution to the internal energy density to be
\be
u(T) = \frac{1}{V}\sum_{\bm p}\,\omega_0 ({\bm p})\, n(\omega_0({\bm p})).
\label{eq:2.2}
\ee
Here $n(x) = 1/(\exp(x/T)-1)$ is the Bose distribution function. $V$ denotes
the system volume, and throughout this paper we use units such that
$k_{\text{B}} = \hbar = 1$. By using Eq.\ (\ref{eq:2.1}) and scaling out the
temperature it is easy to see that the leading helimagnon contribution to the
specific heat, $C=\partial u/\partial T$, is proportional to
$T^2$.\cite{LRO_footnote} Adding the leading Fermi liquid contribution, which
is linear in $T$, the low-temperature specific heat in the helical phase is
\be
C(T \to 0) = \gamma\,T + \gamma_{2}\,T^{2} + O(T^{3}\ln T).
\label{eq:2.3}
\ee
Here $\gamma$ is the usual Fermi-liquid specific heat coefficient, and
$\gamma_2$ is the coefficient of the leading helimagnon contribution. From Eq.\
(\ref{eq:2.2}) we obtain
\be
\gamma_2 = \frac{3\,\zeta (3)}{4\pi}\,\frac{1}{\sqrt{c_{z}c_{\perp}}}\ ,
\label{eq:2.4}
\ee
with $\zeta(x)$ denoting the Riemann zeta-function.
The term of $O(T^3\ln T)$ in Eq.\
(\ref{eq:2.3}) is the usual nonanalytic term in Fermi-liquid theory that also
exists in a nonmagnetic metal.\cite{Baym_Pethick_1991}

It is also interesting to compare the helimagnon contribution to the specific
heat contribution from the Goldstone modes or spin waves in antiferromagnets
and ferromagnets, which have dispersion relations $\omega_0({\bm p})\propto
\vert{\bm p}\,\vert$ and $\omega_0({\bm p})\propto {\bm p}^{\,2}$,
respectively. Equation (\ref{eq:2.2}) yields $C_{\text{afm}} \propto T^3$ and
$C_{\text{fm}}\propto T^{3/2}$ for these two cases.\cite{Kittel_1996} The
helimagnetic case is thus in between the ferromagnetic and antiferromagnetic
cases, as one would expect based on the nature of the respective Goldstone
modes.

\subsection{Quasi-particle relaxation rate, and electrical resistivity}
\label{subsec:II.B}

The other main result of the present paper concerns the temperature dependence
of the quasi-particle relaxation rate and the electrical resistivity due to the
scattering of electrons by helimagnons in a helically ordered magnetic state.

\subsubsection{Quasi-particle relaxation rate}
\label{subsubsec:II.B.1}

\begin{figure}[t,h]
\vskip -0mm
\includegraphics[width=3.0cm]{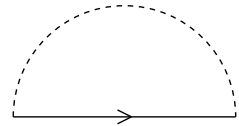}
\caption{The lowest-order electronic self energy $\Sigma$ in terms of the
potential $V$ (dashed line) and the electronic Green function (solid line).}
\label{fig:1}
\end{figure}

To make plausible our result for the quasi-particle relaxation rate, let us
recall the case of electrons with an energy-momentum relation $\epsilon_{\bm
k}$ that are scattered by an effective dynamical potential $V({\bm
p},i\Omega_n)$, with ${\bm p}$ a momentum and $\Omega_n = 2\pi Tn$ a bosonic
Matsubara frequency. The quasi-particle relaxation rate is given by the
imaginary part of the self energy, and to lowest order in the potential the
latter is given by the diagram shown in Fig.\ \ref{fig:1}.
A standard calculation yields the following relaxation rate for a
quasi-particle with wave vector ${\bm k}$ on the Fermi surface,
\be
\frac{1}{\tau_{\bm k}} = \frac{1}{V}\sum_{\bm p} \frac{V''({\bm p}-{\bm
k},\xi_{\bm p})}{\sinh(\xi_{\bm p}/T)}\ ,
\label{eq:2.5}
\ee
where $V''({\bm p},\omega) = \Im V({\bm p},i\Omega_n\rightarrow\omega + i0)$ is
the spectrum of the dynamical potential, and $\xi_{\bm p} = \epsilon_{\bm p} -
\epsilon_{\text{F}}$ with $\epsilon_{\text{F}}$ the Fermi energy.

In the case of a Fermi liquid, the relevant effective potential is the
dynamically screened Coulomb interaction, which has the property $V''({\bm
p}\,,\omega) \propto \omega/p\,v_{\text{F}}$.\cite{Pines_Nozieres_1989} The
integration along the Fermi surface then gives simply a geometrical prefactor,
and the temperature dependence of the relaxation rate is given by the
integration over the modulus of ${\bm p}$,\cite{golden_rule_footnote}
\be
\frac{1}{\tau_{\text{e-e}}} \propto \int_{-\infty}^{\infty}
d\xi\,\frac{\xi}{\sinh(\xi/T)} \propto T^2.
\label{eq:2.6}
\ee

Another well-known example is the scattering of electrons by acoustic phonons,
in which case $V''({\bm p},\omega) \propto cp\,\delta(\omega -
cp)$,\cite{AGD_1963} with $c$ the speed of sound. This leads to
\be
\frac{1}{\tau_{\text{e-ph}}} \propto \frac{1}{V}\sum_{\bm p}
cp\,\frac{\delta(cp - {\bm p}\cdot{\bm k}/m_{\text{e}})}{\sinh(cp/T)} \propto
T^3.
\label{eq:2.7}
\ee
We note that the delta-function contribution to the effective potential
reflects the phonon propagator, whereas the prefactor $cp$ reflects a matrix
element squared that describes the coupling between electrons and
phonons.\cite{Fetter_Walecka_1971} Also, we have ignored a second contribution
to the phonon propagator that leads to the same temperature dependence.

With these preliminary considerations one can make an educated guess about the
helimagnon contribution to the quasi-particle scattering rate. First of all,
the quantity $\xi_{\bm p}$, which measures the distance from the Fermi surface,
needs to be generalized to reflect the existence of two non-spherical Fermi
surfaces split by a Stoner gap. It follows from I Eqs. (4.13), and will be
discussed in more detail in Sec.\ \ref{sec:III} below, that this generalization
reads
\be
\xi_{\bm p} \rightarrow \omega_{1,2}({\bm p}) = \frac{1}{2}\,\left[\xi_{\bm p}
+ \xi_{{\bm p}+{\bm q}} \pm \sqrt{(\xi_{\bm p} - \xi_{{\bm p}+{\bm q}})^2 +
4\lambda^2}\right].
\label{eq:2.8}
\ee
Here we neglect spin-orbit coupling effects which can qualitatively modify the
band structure.\cite{Fischer_Rosch_2004} $\lambda$ is the exchange splitting or
Stoner gap from I Eq. (4.10c). The Fermi surfaces are defined as the loci of
wave vectors ${\bm k}$ with $\omega_{1,2}({\bm k}) = 0$. The inelastic
lifetimes for quasi-particles on either Fermi surface have the same
temperature-dependence, and for definiteness we will concentrate on the Fermi
surface given by $\omega_1({\bm k}) = 0$. The effective potential will be
proportional to the spin susceptibility, which in turn is known from I to be
proportional to the Goldstone propagator. For power-counting purposes, the
spectrum of the latter is adequately represented by (see I Eqs. (4.33))
\be
\chi_{\phi\phi}''({\bm p},\omega) \propto \frac{1}{\omega_0({\bm
p})}\,\delta\left(\omega - \omega_0({\bm p})\right)
\label{eq:2.9}
\ee
Here $\phi$ is the order parameter phase defined in I (see Eqs. (3.4a), (4.17),
(4.28), and (4.32b) in I). In addition, we expect the effective potential to
contain a multiplicative function $g({\bm k},{\bm p})$ that describes the
coupling between electrons and helimagnons, in analogy to the phonon case.
Altogether we expect for the inverse lifetime of a quasi-particle with wave
vector ${\bm k}$ on the 1-Fermi surface due to scattering by helimagnons
\bea
\frac{1}{\tau_{\bm k}} &\propto& \frac{1}{V}\sum_{\bm p} \frac{g({\bm k},{\bm
p})}{\sinh(\omega_1({\bm p})/T)}\,\chi_{\phi\phi}''({\bm p}-{\bm
k},\omega_1({\bm p})).
\nonumber\\
&&\hskip -20pt = \frac{1}{V}\sum_{\bm p} \frac{g({\bm k},{\bm k}+{\bm
p})}{\omega_0({\bm p})\sinh(\omega_0({\bm p})/T)}\,\delta(\omega_1({\bm k}+{\bm
p})-\omega_0({\bm p})).
\nonumber\\
\label{eq:2.10}
\eea
The determination of the function $g$ requires some information about the
coupling mechanism, as in the electron-phonon case. Since $\phi$ is a phase
mode, one expects the physically relevant correlation function to describe the
fluctuations of the gradient of $\phi$, rather than of the phase itself. This
suggests $g({\bm k},{\bm p}) \propto ({\bm k}-{\bm p})^2$. Indeed, we will see
in Sec.\ \ref{sec:III} that for free electrons, $\epsilon_{\bm k} =
k^2/2m_{\text{e}}$, one has $g({\bm k},{\bm k}+{\bm p}) \propto p_z^{\,2}$ for
small deviations ${\bm p}\,$ from the Fermi surface. The ${\bm
p}_{\perp}^{\,2}$ term is absent in this case, as it is in the helimagnon
dispersion relation. At the same time, $\omega_1({\bm k}+{\bm p}) = {\bm
k}_{\perp}\cdot{\bm p}_{\perp}/m_{\text{e}}$. Power counting then shows that
$1/\tau_{\bm k} \propto T^{5/2}$ for generic points on the Fermi surface.
However, for a generic $\epsilon_{\bm k}$ that is consistent with a cubic
symmetry (as is relevant for, e.g., MnSi) one finds $g({\bm k},{\bm k}+{\bm p})
\propto ({\bm k}_{\perp}\cdot{\bm p}_{\perp})^2 + O(k_z^2)$. At the same time,
$\omega_1({\bm k}+{\bm p})$ is modified such that the leading argument of the
delta-function is no longer proportional to ${\bm k}_{\perp}\cdot{\bm
p}_{\perp}$. Since, in a scaling sense, $p_{\perp}^{\,2} \sim p_z \sim
T$,\cite{scaling_footnote} this reduces the power of temperature by one, and
for generic points on the Fermi surface one obtains
\be
\frac{1}{\tau_{\text{e-hm}}} \propto T^{3/2}
\label{eq:2.11}
\ee
for the electron-helimagnon scattering contribution to the quasi-particle
scattering rate. Notice that this is stronger than the standard Fermi-liquid
contribution to the scattering rate, Eq.\ (\ref{eq:2.6}), but not strong enough
to destroy the Fermi liquid.

Equation (\ref{eq:2.10}) also yields a qualitative result for
antiferromagnets,\cite{fm_footnote} whose Goldstone modes have an isotropic
dispersion relation with $\omega_0({\bm p}) \propto \vert{\bm p}\,\vert$. Power
counting with $g({\bm k},{\bm p}) \propto ({\bm k}-{\bm p})^2$ yields
$1/\tau_{\text{e-afm}} \propto T^3$. The temperature dependence is weaker than
in the result for the helimagnet, as one would expect from the nature of the
respective Goldstone modes.

\subsubsection{Electrical conductivity}
\label{subsubsec:II.B.2}

The quasi-particle lifetime is related to, but not the same as, the relevant
time scale for the electrical conductivity $\sigma$, or the resisitivity $\rho
= 1/\sigma$. Technically, the conductivity is given by a four-fermion
correlation function, and vertex corrections enter in addition to the
self-energy contributions that determine the quasi-particle lifetime.
Physically, backscattering events contribute more strongly to the resistivity
than forward scattering events. This leads to a transport scattering rate
$1/\tau_{\text{tr}} \propto \rho$ that is given by Eq. (\ref{eq:2.5}), or the
first line in Eq. (\ref{eq:2.10}), with an extra factor of $({\bm p}-{\bm
k})^2$ in the integrand. In addition, the transport rate gets averaged over the
Fermi surface. In the Coulomb case this changes just the geometric prefactor,
but not the temperature dependence. The Fermi-liquid contribution to the
resistivity thus is\cite{Lifshitz_Pitaevskii_1981}
\be
\rho_{\,\text{e-e}} \propto T^2.
\label{eq:2.12}
\ee
In the case of scattering by a propagating mode, as in the phonon and
helimagnon cases, the momentum is slaved to the frequency by the
delta-function, and hence the temperature dependence does change. In the phonon
case, where the phonon wave number scales as $T$, one obtains the familiar
Bloch-Gr{\"u}neisen result\cite{Ziman_1960}
\be
\rho_{\,\text{e-ph}} \propto T^5.
\label{eq:2.13}
\ee
In the helimagnon case, the smallest (in a scaling sense) additional factor in
the integrand is proportional to $({\bm p}-{\bm k})_{\perp}^2 \sim T$, and the
averaging over the Fermi surface is not important since the behavior at generic
points on the Fermi surface is the leading one. We thus expect
\be
\rho_{\,\text{e-hm}} \propto T^{5/2}.
\label{eq:2.14}
\ee
Adding the Fermi-liquid contribution, we thus obtain for the low-temperature
behavior of the resistivity in the ordered phase of a clean (no impurity
scattering) helimagnet
\be
\rho(T\to 0) = \rho_2\,T^2 + \rho_{5/2}\,T^{5/2} + O(T^3),
\label{eq:2.15}
\ee
with $\rho_2$ and $\rho_{5/2}$ temperature-independent coefficients.

For antiferromagnets the additional factor of $({\bm p}-{\bm k})^2$ in the
integral leads to an additional factor of $T^2$ in the resistivity. The
contributions from the Goldstone modes to the resistivity in this case is thus
expected to be $\rho_{\,\text{afm}} \propto T^5$.\cite{Ueda_1977} Again, the
behavior is weaker than in the helimagnetic case.

The above simple arguments capture the structure of the full theoretical
development in Sec.\ \ref{sec:III}, although they ignore many subtleties that
occur in the actual calculation. The result shows that helimagnon scattering,
while leading to a nonanalytic temperature dependence to the resistivity, is
weaker than the usual electron-electron contribution. Also, the $T^{5/2}$
contribution is one power of $T$ weaker than the observed temperature
dependence in the {\em paramagnetic} phase of
MnSi.\cite{Pfleiderer_Julian_Lonzarich_2001} Nevertheless, it is intriguing
that it is a nonanalytic term and involves a half-integer power of the
temperature. It is also intriguing that the temperature dependence of the
quasi-particle rate in the helimagnetic phase is the same as the observed
transport rate in the paramagnetic phase. We will come back to the possible
relevance of these observations for the paramagnetic phase in Sec.\
\ref{sec:IV}.

\section{Effects of Helimagnons on Observables}
\label{sec:III}

In this section we use the field theory of fluctuations in the helically
ordered phase developed in I to calculate the specific heat and the
quasi-particle relaxation time.

\subsection{Specific heat}
\label{subsec:III.A}

In this subsection we calculate the contribution to the free energy from
Gaussian helimagnon fluctuations and use the result to obtain the leading
fluctuation contribution to the specific heat. We will reproduce the results of
the simple argument given in Sec.\ \ref{sec:II}.

If we ignore all degrees of freedom other than the Goldstone mode $g$, we have
from Eqs.\ I (4.33) the following Gaussian action,
\bse
\label{eqs:3.1}
\be
{\cal A}^{(2)}=\frac{1}{2}\sum_{\bm{p}}\sum_{i\Omega} g({\bm
p},i\Omega)\,\Gamma({\bm p},i\Omega)\,g(-{\bm p},-i\Omega),
\label{eq:3.1a}
\ee
with a vertex function,
\be
\Gamma({\bm p},i\Omega) = -(i\Omega)^2 + \omega_0^2({\bm p}) +
\vert\Omega\vert\,\gamma({\bm p}).
\label{eq:3.1b}
\ee
\ese
Here, and in the remainder of this section, we suppress the integer index on
Matsubara frequencies. $\omega_0({\bm p})$ is the oscillator frequency given by
Eq.\ (\ref{eq:2.1}), and the damping coefficient $\gamma({\bm p})$ is given by
Eq.\ I (4.33c). In this Gaussian approximation, the Goldstone mode contribution
to the grand canonical potential $\Xi$ is
\be
\Xi^{(2)} = -\frac{T}{V}\,\ln \int D[g]\ \exp\left(-{\cal A}^{(2)}[g]\right).
\label{eq:3.2}
\ee
This gives
\be
\Xi^{(2)} = \frac{T}{2V}\sum_{\bm p}\sum_{i\Omega}\ \ln \Gamma({\bm
p},i\Omega).
\label{eq:3.3}
\ee
Neglecting the $i\Omega=0$ term, which does not contribute to the specific
heat, the frequency sum is conveniently performed by using the identity (see
Appendix \ref{app:A})
\bea
T\sum_{i\Omega} \ln\Gamma({\bm p},i\Omega) &=&
\frac{-T}{\pi}\int_0^{\infty}d\omega\,\left[\ln\sinh\frac{\vert\omega\vert}{2T}
- \ln\vert\omega\vert\right]\,
\nonumber\\
&&\times\Im\left(\frac{\partial}{\partial\omega}\,
       \ln\Gamma({\bm p},\omega + i0)\right).
\label{eq:3.4}
\eea
The specific heat at constant volume is obtained from $\Xi$ via the
relation\cite{C_footnote}
\be
C_V = -T\partial^2\Xi/\partial T^2 = \frac{\partial}{\partial T}\,\left[\Xi -
T\,\frac{\partial \Xi}{\partial T}\right].
\label{eq:3.5}
\ee
In the limit of negligible damping, $\gamma({\bm p}) \to 0$, the expression for
$C_V$ can be written, by combining Eqs.\ (\ref{eq:3.4}) and (\ref{eq:3.5}),
\be
C_V(T) = \frac{\partial}{\partial T}\,\frac{1}{V}\sum_{\bm p} \omega_0 ({\bm
p})\,n_{B}(\omega_0 ({\bm p})).
\label{eq:3.6}
\ee
This is identical with the result for $C_V$ obtained from Eq.\ (\ref{eq:2.2}),
and leads to Eq.\ (\ref{eq:2.4}) for the specific heat coefficient. With a
nonzero damping coefficient, the frequency integral cannot be done in closed
form, but the asymptotic low-temperature behavior is still proportional to
$T^2$.\cite{damping_footnote}

The model calculation given in I, which kept modes with wave numbers $k=0$ and
$k=q$, results in a helimagnon dispersion relation (see Eq.\ I (4.33b) and Ref.
\onlinecite{erratum_footnote})
\bse
\label{eqs:3.6'}
\be
\omega_0({\bm k}) =
\lambda\,\frac{q}{3k_{\text{F}}}\sqrt{k_z^2/(2k_{\text{F}})^2 +
\frac{1}{2}\,k_{\perp}^4/(2qk_{\text{F}})^2}.
\label{eq:3.6'a}
\ee
Keeping modes with higher wave numbers changes the dispersion relation to
\be
\omega_0({\bm k}) =
\lambda\,\frac{q}{3k_{\text{F}}}\sqrt{k_z^2/(2k_{\text{F}})^2 +
\frac{3}{8}\,k_{\perp}^4/(2qk_{\text{F}})^2}.
\label{eq:3.6'b}
\ee
\ese
This result, which is analogous to the one obtained for cholesteric liquid
crystals,\cite{Lubensky_1972} leads to the following values for the elastic
constants in (\ref{eq:2.1}),
\bse
\label{eqs:3.7}
\bea
c_z &=& \lambda^2\,q^2/36\,k_{\text{F}}^4,
\label{eq:3.7a}\\
c_{\perp} &=& \lambda^2/96\,k_{\text{F}}^4.
\label{eq:3.7b}
\eea
\ese
Here $k_{\text{F}}$ is the Fermi wave number, and the energy scale $\lambda$ is
the exchange splitting or Stoner gap as defined in I. This results in the
following asymptotic low-temperature behavior of the helimagnon contribution to
the specific heat
\be
C_V^{\text{hm}}(T) = q^3\,A_C\,(T/T_q)^2,
\label{eq:3.8}
\ee
with a coefficient $A_C = \sqrt{6}\zeta(3)/2\pi \approx 0.47$, and $T_q =
\lambda q^2/6k_{\text{F}}^2$. Equation (\ref{eq:3.8}) is valid for temperatures
$T \ll T_q$. We will provide a discussion of this result, as well as a
semi-quantitative analysis, in Sec.\ \ref{sec:IV}.

\subsection{Quasi-particle relaxation time}
\label{subsec:III.B}

In this subsection we calculate the temperature dependence of the
quasi-particle inelastic relaxation time in a helimagnet, i.e., the lifetime of
a free-electron state on the Fermi surface due to scattering by helimagnon
fluctuations. We will calculate this quantity to first order in the helimagnon
susceptibility. We will use the result to determine the low-temperature
behavior of the electrical resistivity.

\subsubsection{Effective action}
\label{subsubsec:III.B.1}

To proceed, we need an electronic action that takes into account the helical
magnetic order and helical magnetic fluctuations. The starting point is an
electronic action of the form
\bse
\label{eqs:3.9}
\be
S[{\bar\psi},\psi] = {\tilde S}_0[{\bar\psi},\psi] + \frac{1}{2}\int dx\,dy\,
n_{\text{s}}^i(x)\,A_{ij}(x-y)\,n_{\text{s}}^j(y).
\label{eq:3.9a}
\ee
Here ${\bar\psi}$ and ${\psi}$ are fermionic (i.e., Grassmann-valued) fields
that depend on space, time, and a spin index $\alpha$; $n_{\text{s}}^i(x) =
{\bar\psi}_{\alpha}(x)\sigma^i_{\alpha\beta}\psi_{\beta}(x)$, where $\sigma^i$
($i=1,2,3$) denotes the Pauli matrices, are the components of the electronic
spin density field ${\bm n}_{\text{s}}(x)$. $x\equiv({\bm x},\tau)$ comprises
the position ${\bm x}$ in real space and the imaginary time $\tau$, and the
related integration measure is given by $\int dx = \int_V d{\bm
x}\int_0^{1/T}d\tau$. ${\tilde S}_0$ contains all parts of the action other
than the spin-triplet interaction. The interaction amplitude $A$ consists of a
point-like Hubbard interaction with an amplitude $\Gamma_{\text{t}}$ and a
chiral part with coupling constant $c$ whose origin has been explained in I,
\be
A_{ij}(x-y) = \delta(x-y)\,\left[\delta_{ij}\,\Gamma_{\text{t}} +
\epsilon_{ijk}\,c\,\Gamma_{\text{t}}\,\partial_k\right].
\label{eq:3.9b}
\ee
\ese
Note that $A$ is static; it depends only on spatial position, and not on
imaginary time.

The general idea is now to replace one of the spin density fields in the last
term in Eq.\ (\ref{eq:3.9a}) by a classical (i.e., c-number valued) field that
represents the effective field seen by the electrons due to the magnetic order.
If one is just interested in incorporating static helical order, one can
implement a mean-field approximation by replacing either one of the spin
density fields in Eq.\ (\ref{eq:3.9a}) by its average value, $n^2 \approx
2n\langle n\rangle - \langle n\rangle^2$. This yields an effective action
\bse
\label{eqs:3.10}
\be
S_0[{\bar\psi},\psi] = {\tilde S}_0[{\bar\psi},\psi] + \int dx\ {\bm H}_0({\bm
x})\cdot{\bm n}_{\text{s}}({\bm x}),
\label{eq:3.10a}
\ee
where
\be
{\bm H}_0({\bm x}) = \Gamma_{\text{t}}\,\langle {\bm n}_{\text{s}}(x)\rangle,
\label{eq:3.10b}
\ee
and we have dropped a constant contribution to the action. The remaining
question regards the equation of state that determines $\langle{\bm
n}_{\text{s}}(x)\rangle$. This has been given in I in a saddle-point
approximation, which yields
\be
{\bm H}_0({\bm x}) = \lambda\,\left(\cos({\bm q}\cdot{\bm x}), \sin({\bm
q}\cdot{\bm x}), 0\right),
\label{eq:3.10c}
\ee
\ese
Here ${\bm q}$ is the pitch vector of the helix, and the amplitude $\lambda$ is
determined by a generalized Stoner equation of state. $S_0$ is the reference
ensemble action given by Eq.\ I (4.7a), which describes noninteracting
electrons (or electrons interacting via a spin-singlet interaction only) in an
effective magnetic field given by the helically modulated magnetization.

More generally, one might ask whether one can replace ${\bm H}_0$ in Eq.\
(\ref{eq:3.10a}) by a {\em fluctuating} classical field ${\bm H}(x) =
\Gamma_{\text{t}}\,{\bm M}(x)$, where ${\bm M}(x)$ represents the spin density
averaged over the quantum mechanical degrees of freedom. For the sake of
presentational simplicity we will explain this procedure for an ordinary
Hubbard interaction ($c=0$ in Eq.\ (\ref{eq:3.9b}) and ${\bm q}=0$ in Eq.\
(\ref{eq:3.10c})). The development for the chiral interaction is exactly
analogous.

Writing ${\bm M}(x) = {\bm H}_0/\Gamma_{\text{t}} + \delta{\bm M}(x)$, and
ignoring a constant contribution to the action, one has
\bse
\label{eqs:3.11}
\be
S[{\bar\psi},\psi,\delta{\bm M}] = S_0[{\bar\psi},\psi] + \Gamma_{\text{t}}
\int dx\ \delta{\bm M}(x)\cdot\delta{\bm n}_{\text{s}}(x),
\label{eq:3.11a}
\ee
with $\delta{\bm n}_{\text{s}}(x) = {\bm n}_{\text{s}}(x) - \langle{\bm
n}_{\text{s}}(x)\rangle$. This needs to be supplemented by an action governing
$\delta{\bm M}$. If the expectation value of ${\bm M}$ is to represent the
exact magnetization, this action must be
\be
{\cal A}[\delta{\bm M}] = -\frac{1}{2}\int dx\,dy\ \delta
M_i(x)\,\left(\chi_{\text{s}}^{-1}\right)_{ij}(x,y)\,\delta M_j(y),
\label{eq:3.11b}
\ee
\ese
where $\chi_{\text{s}}$ is the spin susceptibility of the system. A purely
electronic effective action is now obtained by integrating out the fluctuations
$\delta{\bm M}$. The Gaussian integral yields
\be
S_{\text{eff}}[{\bar\psi},\psi] = S_0[{\bar\psi},\psi] +
\frac{\Gamma_{\text{t}}^2}{2}\int dx\, dy\ \delta n_{\text{s}}^i(x)\,
\chi_{\text{s}}^{ij}(x,y)\,\delta n_{\text{s}}^j(y).
\label{eq:3.12}
\ee
With the full chiral interaction, Eq.\ (\ref{eq:3.9b}), one obtains formally
the same result. The only reservation is that in the chiral case the
correlation function $\langle\delta M\,\delta M\rangle = \chi_{\text{s}}$ is,
strictly speaking, not the physical spin susceptibility. However, its leading
hydrodynamic contribution is the same as that of the latter, as was shown in I.
The result (\ref{eq:3.12}) thus consists of the reference ensemble described by
$S_0$, whose Green function has been determined in I, and an effective
interaction given by the spin susceptibility in the helically ordered phase.
The latter has been calculated in I in a Gaussian approximation. Notice that
the effective potential depends on imaginary time or frequency.

\begin{figure}[t,h]
\vskip -0mm
\includegraphics[width=6.0cm]{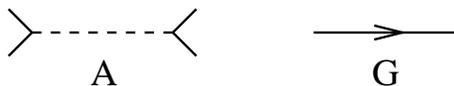}
\caption{Diagrammatic representation of the bare spin-triplet interaction $A$
and the reference ensemble Green function $G$.}
\label{fig:2}
\end{figure}

\begin{figure}[t,h]
\vskip -0mm
\includegraphics[width=8.0cm]{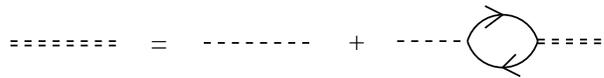}
\caption{Effective spin-triplet interaction resulting from a ladder
resummation.}
\label{fig:3}
\end{figure}

It is obvious that the above heuristic considerations represent, in a
diagrammatic language, some kind of infinite resummation, which means that the
resulting effective action is valid only in conjunction with certain
constraints. To clarify this point it is useful to rewrite the starting point,
Eq.\ (\ref{eq:3.9a}), in the form
\be
S[{\bar\psi},\psi] = S_0[{\bar\psi},\psi] + \frac{1}{2}\int dx\,dy\,\delta
n_{\text{s}}^i(x)\,A_{ij}(x-y)\,\delta n_{\text{s}}^j(y).
\label{eq:3.13}
\ee
Now consider the bare spin-triplet interaction $A$, Fig.\ \ref{fig:2}, and its
renormalization by the ladder resummation shown in Fig.\ \ref{fig:3}.
The result of this resummation is an effective interaction of the form given in
Eq.\ (\ref{eq:3.12}), with a Gaussian approximation for $\chi_{\text{s}}$. The
constraint that must be used in conjunction with the effective action
(\ref{eq:3.12}) is now obvious: The effective interaction must not be used in
ways that constitute renormalizations of $\chi_{\text{s}}$, as doing so would
result in double counting. In particular, it is safe to use Eq.\
(\ref{eq:3.12}) in any perturbative calculation to linear order in
$\chi_{\text{s}}$. We will now proceed and use it to calculate the
quasi-particle inelastic lifetime to that order.

\subsubsection{Electronic self energy and Green function}
\label{subsubsec:III.B.2}

\begin{figure}[h,t]
\includegraphics[width=6.0cm]{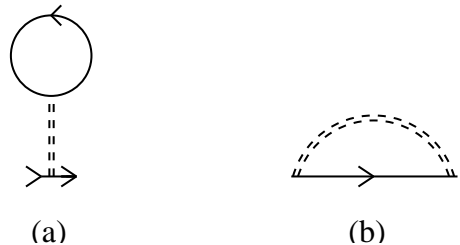}
\caption{The direct (a) and exchange (b) contributions to the single-particle
self energy.}
\label{fig:4}
\end{figure}

We consider the self energy $\Sigma$ of the single-particle Green function to
first order in the perturbing potential $\Gamma_{\text{t}}^2\,\chi_{\text{s}}
\equiv {\tilde\chi}_{\text{s}}$. There are two self-energy diagrams to this
order, namely, the direct or Hartree and the exchange or Fock contributions
shown in Figs.\ \ref{fig:4}(a) and \ref{fig:4}(b), respectively.
Both self-energy contributions are nondiagonal in both spin and momentum space.
It is readily seen that the direct contribution is purely real, and hence does
not contribute to the scattering rate. The exchange contribution is given by,
\bea
\Sigma^{\,\text{ex}}_{\,{\bm k}{\bm p}}(i\omega) &=& \frac{1}{V} \sum_{{\bm
k}',{\bm p}'} T\sum_{i\Omega} \sigma^{i}\,G_{{\bm k}'+{\bm k},{\bm p}'+{\bm
p}}(i\Omega + i\omega)\,\sigma^{j}\,
\nonumber\\
&& \hskip 40pt \times {\tilde\chi}_{\text{s}}^{ij}({\bm k}',{\bm p}';i\Omega).
\label{eq:3.14}
\eea
Here $i\omega$ denotes a fermionic Matsubara frequency, and $G_{{\bm k}{\bm
p}}(i\omega)$ is the single-particle Green function of the reference ensemble
which is given explicitly by Eqs.\ I (4.13). All components of the self energy
can be obtained from Eq.\ (\ref{eq:3.14}) by means of straightforward, albeit
lengthy, calculations.

We now consider the Dyson equation for the renormalized Green function ${\cal
G}$, which reads
\be
{\cal G}^{-1}_{{\bm k} {\bm p}}(i\omega) = G^{-1}_{{\bm k} {\bm p}}(i\omega) -
\Sigma_{\,{\bm k} {\bm p}}(i\omega).
\label{eq:3.15}
\ee
From Eqs.\ I (4.13) for $G$ and Eq.\ (\ref{eq:3.14}) it follows that $\Sigma$
has a structure very similar to that of $G^{-1}$,
\bea
\Sigma_{\,{\bm k}{\bm p}}(i\omega) &=& \delta_{{\bm k}{\bm
p}}\,\left[\sigma_{+-}\,\Sigma_{++}({\bm k},i\omega) +
\sigma_{-+}\,\Sigma_{--}({\bm k},i\omega)\right]
\nonumber\\
&& +\, \delta_{{\bm k}+{\bm q},{\bm p}}\,\sigma_+\,\Sigma_{+-}({\bm k},i\omega)
\nonumber\\
&& +\, \delta_{{\bm k}-{\bm q},{\bm p}}\,\sigma_-\,\Sigma_{-+}({\bm
k},i\omega).
\label{eq:3.16}
\eea
Here the notation is the same as in Eqs.\ I (4.13). In particular,
$\sigma_{\pm} = (\sigma_1 \pm i\sigma_2)/2$, $\sigma_{+-} = \sigma_+ \sigma_-$,
and $\sigma_{-+} = \sigma_- \sigma_+$, with $\sigma_{1,2}$ Pauli matrices. From
this expression it follows in turn that ${\cal G}$ has the same structure as
$G$,
\bse
\label{eqs:3.17}
\bea
{\cal G}_{{\bm k}{\bm p}}(i\omega) &=& \delta_{{\bm k}{\bm
p}}\,\bigl[\sigma_{+-}\,A_+({\bm k},i\omega) + \sigma_{-+}\,A_-({\bm
k},i\omega)\bigr]
\nonumber\\
&& +\, \delta_{{\bm k}+{\bm q},{\bm p}}\,\sigma_+\,B_+({\bm k},i\omega)
\nonumber\\
&& +\, \delta_{{\bm k}-{\bm q},{\bm p}}\,\sigma_-\,B_-({\bm k},i\omega).
\label{eq:3.17a}
\eea
Here
\bea
A_{\pm}({\bm k},i\omega) &=& \frac{g^{-1}_{\mp}({\bm k}\pm {\bm q}, i\omega)}
{g^{-1}_{\mp}({\bm k}\pm{\bm q},i\omega)\,g^{-1}_{\pm}({\bm k},i\omega) -
\left(\lambda_{\pm}({\bm k},i\omega)\right)^2} \nonumber\\
\label{eq:3.17b}\\
B_{\pm}({\bm k},i\omega) &=& \frac{-\lambda_{\pm}({\bm k}, i\omega)}
{g^{-1}_{\mp}({\bm k}\pm{\bm q},i\omega)\,g^{-1}_{\pm}({\bm k},i\omega) -
\left(\lambda_{\pm}({\bm k},i\omega)\right)^2} \nonumber\\
\label{eq:3.17c}
\eea
with
\be
g^{-1}_{\pm}({\bm k},i\omega) = G^{-1}({\bm k},i\omega)
   - \Sigma_{\pm\pm}({\bm k},i\omega),
\label{eq:3.17d}
\ee
\be
\lambda_{\pm}({\bm k},i\omega) = \lambda - \Sigma_{\pm\mp}({\bm k},i\omega).
\label{eq:3.17e}
\ee
\ese
These expressions constitute an exact inverse of Eq.\ (\ref{eq:3.15}), as can
easily be checked by a direct multiplication.

\subsubsection{Quasi-particle relaxation time}
\label{subsubsec:III.B.3}

The quasi-particle relaxation time is determined by the imaginary parts of the
poles of the Green function ${\cal G}$, Eqs.\ (\ref{eqs:3.17}). For a vanishing
self energy there are poles at
\be
\omega_{1,2}^{\pm}({\bm k}) = \frac{1}{2}\,\left( \xi_{\bm k} + \xi_{{\bm
k}\pm{\bm q}} \pm \sqrt{\left(\xi_{\bm k} - \xi_{{\bm k}\pm{\bm q}}\right)^2 +
4\lambda^2}\right), \nonumber
\ee
with $\xi_{\bm k}$ as defined after Eq.\ (\ref{eq:2.5}). The poles with
different signs of the square root reflect the Stoner splitting of the Fermi
surface into two sheets. On a given sheet, $\omega^+$ and $\omega^-$ are
related by $\omega_i^-({\bm k}+{\bm q}) = \omega_i^+({\bm k})$ ($i=1,2$). All
poles can thus be expressed in terms of
\be
\omega_{1,2}({\bm k}) = \frac{1}{2}\,\left( \xi_{\bm k} + \xi_{{\bm k}+{\bm q}}
\pm \sqrt{\left(\xi_{\bm k} - \xi_{{\bm k}+{\bm q}}\right)^2 +
4\lambda^2}\right).
\label{eq:3.18}
\ee
Both of these resonance frequencies are real, reflecting the fact that the
quasi-particles are infinitely long lived to zeroth order in the effective
potential $\chi_{\text{s}}$. To first order in $\chi_{\text{s}}$ the resonance
frequencies acquire an imaginary part, corresponding to a finite relaxation
time $\tau({\bm k})$, in addition to a shift of the real part. For
definiteness, we consider the resonance at $\omega_1({\bm k})$; the relaxation
time on the other sheet has the same temperature dependence. We find
\begin{widetext}
\bea
\frac{1}{\tau({\bm k})} &=& {\text{Im}}\left\{\Sigma_{++}({\bm k},z) +
\Sigma_{--}({\bm k}+{\bm q},z) + \frac{(\xi_{\bm k} - \xi_{{\bm k}+{\bm
q}})}{\left[\left(\xi_{\bm k} - \xi_{{\bm k}+{\bm q}}\right)^2 +
4\lambda^2\right]^{1/2}}\,\left[\Sigma_{++}({\bm k},z) - \Sigma_{--}({\bm
k}+{\bm q},z)\right]\right.
\nonumber\\
&& \hskip 20pt \left. - \frac{2\lambda}{\left[\left(\xi_{\bm k}-\xi_{{\bm
k}+{\bm q}}\right)^2 + 4\lambda^2\right]^{1/2}} \, \left[\Sigma_{+-}({\bm k},z)
+ \Sigma_{-+}({\bm k}+{\bm q},z)\right]\right\}\ .
\label{eq:3.19}
\eea
\end{widetext}
Here the frequency $z$ is given by the `on-shell condition' $z = \omega_1({\bm
k}) + i0$.

Of particular interest is the relaxation rate for quasi-particles on the Fermi
surface, which is defined by
\bse
\label{eqs:3.20}
\bea
\omega_1({\bm k}) &=& \frac{1}{2}\,\left( \xi_{\bm k} + \xi_{{\bm k}+{\bm q}} +
\left[\left(\xi_{\bm k} - \xi_{{\bm k}+{\bm q}}\right)^2 +
4\lambda^2\right]^{1/2}\right)
\nonumber\\
&=& 0.
\label{eq:3.20a}
\eea
For $\lambda = {\bm q} = 0$ this reduces to the usual definition of the Fermi
surface, $\xi_{\bm k}=0$. In the general case it implies in particular
\be
\xi_{\bm k}\,\xi_{{\bm k}+{\bm q}} = \lambda^2.
\label{eq:3.20b}
\ee
\ese
With this condition, we use Eq.\ (\ref{eq:3.14}) in Eq.\ (\ref{eq:3.19}), and
the results of I for $\chi_{\text{s}}$, keeping only the leading hydrodynamic
contributions to the latter. Performing the frequency summations we find for
the relaxation rate of a quasi-particle on the Fermi surface
\bse
\label{eqs:3.21}
\bea
\frac{1}{\tau_{\text{F}}({\bm k})} &=& \frac{-2\lambda^2}{\left(\xi_{\bm k} +
\xi_{{\bm k}+{\bm q}}\right)^2}\,\frac{1}{V}\sum_{\bm p} \ \frac{g({\bm k},{\bm
p})}{\sinh\left(\omega_1({\bm p})/T\right)}\
\nonumber\\
&& \hskip 40pt \times \chi_{\phi\phi}''\left({\bm p}-{\bm k},\omega_1({\bm
p})\right).
\label{eq:3.21a}
\eea
This expression is valid to determine the leading low-temperature behavior of
$1/\tau_{\text{F}}$ only. We have anticipated the fact that the dominant
contribution to the latter comes from momenta ${\bm p} \approx {\bm k}$, and
accordingly have replaced ${\bm p}$ by ${\bm k}$ in all contributions to the
integrand that have a finite and nonzero limit as ${\bm p} \to {\bm k}$. Here
$\chi_{\phi\phi}''$ is the spectrum of the susceptibility of the phase
fluctuation variable $\phi$ that was introduced in I. Its leading hydrodynamic
part is also proportional to the Goldstone mode. From I (4.33a) or,
alternatively, from I (4.29a) and (4.40),\cite{erratum_footnote} we find for
the phase susceptibility
\be
\chi_{\phi\phi}({\bm p},i\Omega) = \frac{1}{2N_{\text{F}}}\,\frac{q^2}{3
k_{\text{F}}^2}\,\frac{1}{\omega_0^2({\bm p}) - (i\Omega)^2}\ ,
\label{eq:3.21b}
\ee
where $\omega_0({\bm p})$ is the helimagnon resonance frequency, see Eqs.\
(\ref{eq:2.1}) and I (4.33b). The spectrum is thus
\bea
\chi_{\phi\phi}''({\bm p},\omega) &=& \frac{1}{2N_{\text{F}}}\,\frac{q^2}{3
k_{\text{F}}^2}\,\frac{\pi}{2\omega_0({\bm p})}\bigl[\delta\left(\omega -
\omega_0({\bm p})\right)
\nonumber\\
&& \hskip 50pt - \delta\left(\omega + \omega_0({\bm p})\right)\bigr].
\label{eq:3.21c}
\eea
The function $g$ is given by
\bea
g({\bm k},{\bm p}) &=& \omega_1({\bm p})\left(\xi_{\bm k} + \xi_{{\bm k}+{\bm
q}}\right) + \xi_{\bm k}\left(\xi_{{\bm k}+{\bm q}} - \xi_{{\bm p}+{\bm
q}}\right)\hskip 30pt
\nonumber\\
&&\hskip 78pt + \xi_{{\bm k}+{\bm q}}\left(\xi_{\bm k} - \xi_{\bm p}\right)\ .
\label{eq:3.21d}
\eea
\ese
Notice that $g({\bm k},{\bm k})=0$, that $\chi''({\bm p},\omega)$ is soft at
${\bm p}=0$ and $\omega = \omega_1({\bm k})=0$, and that Eq.\ (\ref{eq:3.21a})
has indeed the form of the educated guess, Eq.\ (\ref{eq:2.10}).

For later reference we also introduce the `off-shell' rate $1/\tau({\bm
k},\epsilon)$ obtained by putting $z=\epsilon + i0$ in the self energies in
Eq.\ (\ref{eq:3.19}). With ${\bm k}$ on either Fermi surface one finds
\begin{widetext}
\bse
\label{eqs:3.22}
\be
\frac{1}{\tau_i({\bm k},\epsilon)} = \frac{-2\pi\lambda^2}{\left(\xi_{\bm k} +
\xi_{{\bm k}+{\bm q}}\right)^2} \int \frac{du}{\pi}\,\left[n(u) + f(\epsilon +
u)\right]\, \frac{1}{V}\sum_{\bm p} g_i({\bm k},{\bm
p}\,;\epsilon)\,\chi_{\phi\phi}''({\bm p}-{\bm
k},u)\,\delta\left(u+\epsilon-\omega_i({\bm p})\right),\quad(i=1,2)\ ,
\label{eq:3.22a}
\ee
Here $\omega_i({\bm p})$ is given by Eq.\ (\ref{eq:3.18}), and
\be
g_i({\bm k},{\bm p}\,;\epsilon) = \left(\omega_i({\bm p}) - \xi_{\bm
p}\right)(\xi_{{\bm k}+{\bm q}}-\epsilon) - \left(\omega_i({\bm p}) - \xi_{{\bm
p}+{\bm q}}\right)\,\frac{\xi_{\bm k}\xi_{{\bm k}+{\bm q}}}{\epsilon -
\xi_{{\bm k}+{\bm q}}} + 2\xi_{\bm k}\,\xi_{{\bm k}+{\bm q}}
\label{eq:3.22b}
\ee
\ese
\end{widetext}
is a generalization of Eq.\ (\ref{eq:3.21d}), with the property $g_1({\bm
k},{\bm p}\,;\epsilon=0) = g({\bm k},{\bm p})$.


To evaluate the final integral we consider the limit $\epsilon_{\text{F}} \gg
\lambda \gg qv_{\text{F}}$, as we did in I. The result will obviously depend on
the direction of ${\bm k}$. It also depends on the functional form electronic
energy momentum relation $\epsilon_{\bm k}$.

\paragraph{Isotropic energy-momentum relation}
\label{par:III.B.3.a}

Let us first consider a nearly-free electron approximation, with
\be
\epsilon_{\bm k} = {\bm k}^2/2m_{\text{e}},
\label{eq:3.23}
\ee
with $m_{\text{e}}$ the effective mass of the electrons. Then $g({\bm k},{\bm
k}+{\bm p}) \propto p_z^2 + O(p_z^3)$, and $\omega_1({\bm k}+{\bm p}) = {\bm
k}_{\perp}\cdot{\bm p}_{\perp}/m_{\text{e}} + O(k_z p_z, k_{\perp}^2)$.

For ${\bm k}_{\perp}=0$ the leading result as $T\to 0$ is
\bse
\label{eqs:3.24}
\bea
\frac{1}{\tau (\bm{k})} &=& \pi^3 \sqrt{6}\ \frac{\lambda
q^4}{m_{\text{e}}^2}\,\frac{k_{\text{F}}^3}{k_z^3}\,\frac{T^2}{(\xi_{\bm k} +
\xi_{{\bm k}+{\bm q}})^4}
\nonumber\\
&\propto& \lambda\left(\frac{q}{k_{\text{F}}}\right)^8
\left(\frac{\epsilon_{\text{F}}}{\lambda}\right)^2
\left(\frac{T}{T_q}\right)^2,
\label{eq:3.24a}
\eea
with $T_q$ the temperature scale defined after Eq.\ (\ref{eq:3.8}). For ${\bm
k}_{\perp}\neq 0$ the asymptotic result is
\bea
\frac{1}{\tau (\bm{k})} &=& A_{\tau}\,\frac{\lambda^{1/2} q k_{\text{F}}^3}
{m_{e}^{2}}\,\frac{k_{\text{F}}}{k_{\perp}}\,\frac{T^{5/2}}{(\xi
_{\bm{k}}+\xi_{{\bm k}+{\bm q}})^4}
\nonumber\\
&\propto& \lambda\,\left(\frac{q}{k_{\text{F}}}\right)^6
\left(\frac{\epsilon_{\text{F}}}{\lambda}\right)^2
\left(\frac{T}{T_q}\right)^{5/2}.
\label{eq:3.24b}
\eea
\ese
with $A_{\tau}\approx 289$. In the second lines in Eqs.\ (\ref{eq:3.24a}) and
(\ref{eq:3.24b}) we have assumed a generic value of ${\bm k}$, and have
neglected all numerical prefactors as well as terms small in
$qv_{\text{F}}/\lambda$ for clarity. The crossover between these two types of
behavior occurs at a temperature proportional to $T_{\times} =
\lambda\,(q/k_{{\text{F}}})^6(k_{\perp }/k_z)^2$. It is easy to check that
adding an isotropic quartic term, proportional to $({\bm k}^2)^2$, to
$\epsilon_{\bm k}$ does not change the powers of the temperature in these
results.

\paragraph{Cubic energy-momentum relation}
\label{par:III.B.3.b}

In an actual metal, the underlying lattice structure causes the electronic
energy-momentum relation to be anisotropic. In the case of a cubic lattice, as
in MnSi, any terms consistent with cubic symmetry are allowed. For instance, to
quartic order in ${\bm k}$ the following function is allowed,
\be
\epsilon_{\bm k} = {\bm k}^2/2m_{\text{e}} +
\frac{\nu}{2m_{\text{e}}k_{\text{F}}^2}\, (k_x^2 k_y^2 + k_y^2 k_z^2 + k_z^2
k_x^2),
\label{eq:3.25}
\ee
with $\nu$ a dimensionless measure of deviations from a nearly-free electron
model. Generically, once expects $\nu = O(1)$. Other quartic terms that are
consistent with a cubic symmetry, e.g., the cubic anisotropy $k_x^4 + k_y^4 +
k_z^4$, can be obtained by adding an isotropic $({\bm k}^2)^2$ term to Eq.\
(\ref{eq:3.25}). For $\nu\neq 0$ the asymptotic behavior of $g$ is changed
compared to the nearly-free electron case,
\bea
g({\bm k},{\bm k}+{\bm p}) &=&
-\left(\frac{2\nu}{2m_{\text{e}}k_{\text{F}}^2}\right)^2
   \,\frac{\lambda^2\,q^2\,(2k_z+q)^2}{(\xi_{\bm k} + \xi_{{\bm k}+{\bm q}})^2}\,
   ({\bm k}_{\perp}\cdot{\bm p}_{\perp})^2
\nonumber\\
&&   + O(p_{\perp}p_z)
\label{eq:3.26}
\eea
and for ${\bm q}$ in $(0,0,1)$-direction one has
\bse
\label{eqs:3.27}
\bea
\omega_1({\bm k}+{\bm p}) &=& ({\bm k}_{\perp}\cdot{\bm
p}_{\perp})/\mu_{\text{e}}
\nonumber\\
&&+ \frac{2\nu}{2m_{\text{e}}k_{{\text F}}^2}\,(k_y^2 + k_z^2 + k_zq +
q^2/2)k_x p_x
\nonumber\\
&&+ \frac{2\nu}{2m_{\text{e}}k_{{\text F}}^2}\,(k_x^2 + k_z^2 + k_zq +
q^2/2)k_y p_y
\nonumber\\
&&+ O(p_{\perp}^2,p_z),
\label{eq:3.27a}
\eea
where
\be
1/\mu_{\text{e}} = 1/m_{\text{e}} + \nu\,\frac{q(2k_z +
q)}{2m_{\text{e}}k_{\text{F}}^2}\,\frac{\xi_{\bm k} - \xi_{{\bm k}+{\bm
q}}}{\xi_{\bm k} + \xi_{{\bm k}+{\bm q}}}.
\label{eq:3.27b}
\ee
\ese
If we define
\bse
\label{eqs:3.28}
\be
A_{x,y} = 1 + \frac{\nu\mu_{\text{e}}}{m_{\text{e}}k_{\text{F}}^2}\,(k_{y,x}^2
+ k_z^2 + k_zq + q^2/2),
\label{eq:3.28a}
\ee
and
\be
C_{\bm k} =
\frac{B\,\nu^4}{k_{\text{F}}^5}\,\frac{(2k_z+q)^2}{k_{\text{F}}^2}\,
\frac{q^3k_{\text{F}}\mu_{\text{e}}^3}{m_{\text{e}}^5}\,
\frac{\lambda^3}{(\xi_{\bm k} + \xi_{{\bm k}+{\bm q}})^4},
\label{eq:3.28b}
\ee
with
\bea
B &=& \frac{48 \sqrt{3}}{2^{1/4}3^{3/4}} \int_0^{\infty} dx\,dz\
\frac{x^2}{\sqrt{z^2 + x^4}}\,\frac{1}{\sinh{\sqrt{z^2 + x^4}}} \nonumber\\
&=& 54.99\ldots
\label{eq:3.28c}
\eea
the evaluation of Eqs.\ (\ref{eqs:3.21}) now yields
\bea
\frac{1}{\tau({\bm k})} &=& C_{\bm k}\,\frac{k_x^2 k_y^2 (k_x^2 -
k_y^2)^2}{(k_x^2 A_x^2 + k_y^2
A_y^2)^{3/2}}\,\left(\frac{T}{\lambda}\right)^{3/2}
\nonumber\\
&\propto& \nu^4 \lambda\left(\frac{q}{k_{\text{F}}}\right)^6
\left(\frac{\epsilon_{\text{F}}}{\lambda}\right)^2
\left(\frac{T}{T_q}\right)^{3/2}.
\label{eq:3.28d}
\eea
\ese

In a cubic system, the quasi-particle relaxation rate thus generically shows a
$T^{3/2}$ behavior, except on four special lines in ${\bm k}$-space, $k_x=0$,
$k_y=0$, and $k_x = \pm k_y$, where the prefactor of the $T^{3/2}$ vanishes and
the behavior is $T^2$. These results are valid for ${\bm q}$ in
$(0,0,1)$-direction. The corresponding expressions for ${\bm q}$ in
$(1,1,1)$-direction, which is relevant for MnSi,\cite{q_footnote} can be
obtained by a corresponding rotation of ${\bm k}$. That is, $(k_x,k_y,k_z)$ in
the above expressions should be replaced by
\bea
k_x &\rightarrow& (k_x - \sqrt{3} k_y + \sqrt{2} k_z)/\sqrt{6},
\nonumber\\
k_y &\rightarrow& (k_x + \sqrt{3} k_y + \sqrt{2} k_z)/\sqrt{6},
\nonumber\\
k_z &\rightarrow& (-2k_x + \sqrt{2} k_z)/\sqrt{6}.
\label{eq:3.29}
\eea

It needs to be stressed that the $T^{5/2}$ and $T^{3/2}$ terms due to the
helimagnons, as well as the Fermi-liquid $T^2$ term, all contribute to the
quasi-particle relaxation rate, and which of these contributions dominate in a
given temperature regime is a quantitative question. The same is true for the
different contributions to the specific heat. In Sec.\ \ref{subsec:IV.A} we
will give estimates for parameter values that are appropriate for MnSi.

\subsection{Resistivity}
\label{subsec:III.C}

\begin{figure}[t,h]
\vskip -0mm
\includegraphics[width=8.0cm]{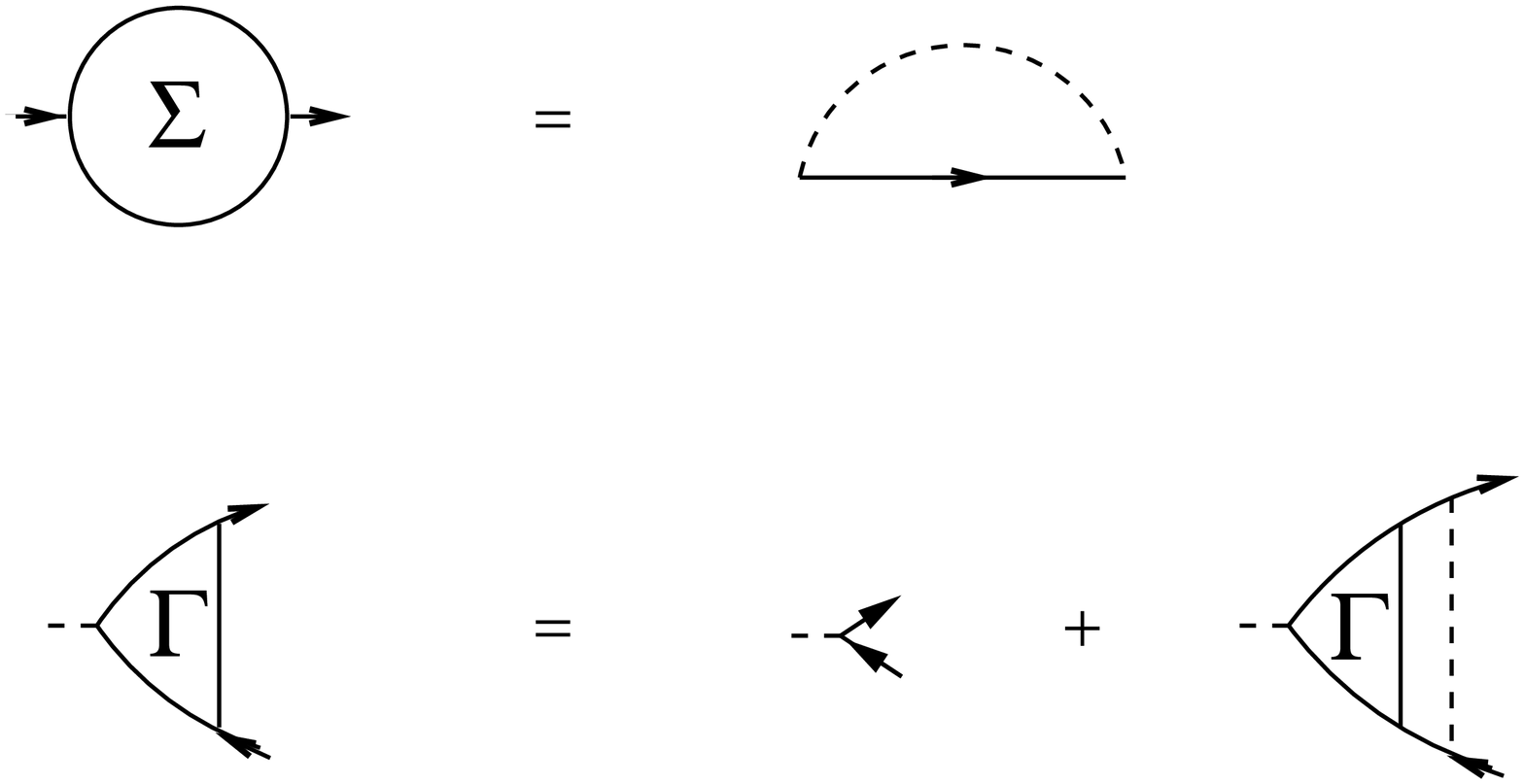}
\caption{Consistent approximation for the self energy $\Sigma$ and the vertex
function ${\bm\Gamma}$.}
\label{fig:5}
\end{figure}

The electrical resistivity can be obtained as the inverse of the conductivity,
which in turn is given by the Kubo formula
\bse
\label{eqs:3.30}
\be
\sigma_{ij}(i\Omega) = \frac{i}{i\Omega}\,\left[\pi_{ij}(i\Omega) -
\pi_{ij}(i\Omega=0)\right].
\label{eq:3.30a}
\ee
Here
\bea
\pi_{ij}(i\Omega) &=& -e^2\,T\sum_{n_1,n_2}\frac{1}{V} \sum_{{\bm k},{\bm p}}
v_i({\bm k})\,v_j({\bm p})\,\hskip 0pt
\nonumber\\
&&\hskip -20pt \times \left\langle{\bar\psi}_{n_1,\sigma}({\bm
k})\,\psi_{n_1+n,\sigma}({\bm k})\,{\bar\psi}_{n_2,\sigma'}({\bm
p})\,\psi_{n_2- n,\sigma'}({\bm p})\right\rangle. \nonumber\\
\label{eq:3.30b}
\eea
\ese
is the current-current susceptibility tensor or polarization function, with
${\bm v}({\bm k}) = \partial\epsilon_{\bm k}/\partial{\bm k}$. The average,
denoted by $\langle\ldots\rangle$, is to be performed with the effective action
given in Eq.\ (\ref{eq:3.12}). The simplest approximation to $\pi$ is a
factorization of the four-point correlation function in Eq.\ (\ref{eq:3.28b})
into two Green functions. The difference between this simple approximation and
the full polarization function is customarily expressed in terms of a vector
vertex function ${\bm\Gamma}$ with components $_i\Gamma$,
\bea
\pi_{ij}(i\Omega) &=& -e^2\,T\sum_{i\omega}\frac{1}{V} \sum_{{\bm k},{\bm
p}}\sum_{\alpha\beta} {_i\Gamma}^{\alpha\beta}_{{\bm k}{\bm p}}(i\omega,i\omega
- i\Omega)
\nonumber\\
&&\hskip -40pt \times \sum_{{\bm k}'}\sum_{\sigma} iv_j({\bm k}')\,{\cal
G}_{{\bm k}{\bm k}'}^{\alpha\sigma}(i\omega)\, {\cal G}_{{\bm k}'{\bm
p}}^{\sigma\beta}(i\omega - i\Omega),
\label{eq:3.31}
\eea
with ${\cal G}$ from Eq.\ (\ref{eq:3.15}). It is well known that care must be
taken to use consistent approximations for the self energy $\Sigma$ that
defines ${\cal G}$ and the vertex function.\cite{Baym_Kadanoff_1961,
Kadanoff_Baym_1962} The simplest combination that fulfills the consistency
requirement, which is equivalent to the Boltzmann equation for the
conductivity, is a self-consistent Born approximation for the self energy,
\bse
\label{eqs:3.32}
\bea
\Sigma_{{\bm k}{\bm p}}^{\alpha\beta}(i\omega) &=& T\sum_{i\Omega}
\frac{1}{V}\sum_{{\bm k}'{\bm p}'} \sum_{\alpha'\beta'} {\cal G}_{{\bm k}'{\bm
p}'}^{\alpha'\beta'}(i\omega + i\Omega)\, \nonumber\\
&&\times V_{\alpha\alpha'\!,\,\beta'\beta}({\bm k}'-{\bm k},{\bm p}\,'-{\bm
p}\,;i\Omega),
\label{eq:3.32a}
\eea
and a ladder approximation for the vertex function,
\begin{widetext}
\bea
{\bm\Gamma}_{{\bm k}{\bm p}}^{\alpha\beta}(i\omega,i\omega - i\Omega) &=&
\delta_{{\bm k}{\bm p}}\,\delta_{\alpha\beta}\,iv({\bm k}) + T\sum_{i\Omega'}
\frac{1}{V}\sum_{{\bm k}'\!,\,{\bm k}''\!,\,{\bm p}'\!,\,{\bm p}''}
{\bm\Gamma}_{{\bm k}''{\bm p}''}^{\alpha''\beta''}(i\omega + i\Omega',i\omega -
i\Omega + i\Omega')\,{\cal G}_{{\bm k}''{\bm k}'}^{\alpha''\alpha'}(i\omega +
i\Omega')\hskip 60pt
\nonumber\\
&&\times\,{\cal G}_{{\bm p}'{\bm k}''}^{\beta'\alpha''}(i\omega - i\Omega +
i\Omega')\,\frac{1}{2}\,\left[V_{\alpha'\alpha\,,\beta\beta'}({\bm k}-{\bm
k}',{\bm p} - {\bm p}\,';i\Omega') + V_{\beta\beta'\!,\,\alpha'\alpha}({\bm
p}\,'-{\bm p}\,,{\bm k}' - {\bm k};i\Omega')\right].
\label{eq:3.32b}
\eea
\end{widetext}
Here $V$ is the potential in the effective action,
\be
V_{\alpha\beta,\gamma\delta}({\bm k},{\bm p}\,;i\Omega) = \sum_{ij}
\sigma^i_{\alpha\beta}\,\tilde{\chi}_{\text{s}}^{ij} ({\bm k},{\bm
p}\,;i\Omega)\,\sigma^j_{\gamma\delta},
\label{eq:3.32c}
\ee
\ese
with ${\tilde\chi}_{\text{s}}$ the spin susceptibility times
$\Gamma_{\text{t}}^2$ as used in the calculation of the quasi-particle
relaxation time in Sec.\ \ref{subsec:III.B}, see Eq.\ (\ref{eq:3.14}).
Equations (\ref{eqs:3.30}) are represented diagrammatically in Fig.
\ref{fig:5}.
With the results for the spin susceptibility given in I (4.41) - (4.43) we find
that only four matrix elements of the potential $V$ are nonzero. They can be
expressed in terms of the phase susceptibility, Eq.\ (\ref{eq:3.21b}),
\bse
\label{eqs:3.33}
\bea
V_{12,21}({\bm k},{\bm p};i\Omega) &=& \delta_{{\bm k}{\bm
p}}\,\lambda^2\,\chi_{\phi\phi}({\bm k}-{\bm q},i\Omega),
\label{eq:3.33a}\\
V_{21,21}({\bm k},{\bm p};i\Omega) &=& \delta_{{\bm k}{\bm
p}}\,\lambda^2\,\chi_{\phi\phi}({\bm k}+{\bm q},i\Omega),
\label{eq:3.33b}\\
V_{12,12}({\bm k},{\bm p};i\Omega) &=& -\delta_{{\bm k}-2{\bm q},{\bm
p}}\,\lambda^2\,\chi_{\phi\phi}({\bm k}-{\bm q},i\Omega), \nonumber\\
\label{eq:3.33c}\\
V_{21,21}({\bm k},{\bm p};i\Omega) &=& -\delta_{{\bm k}+2{\bm q},{\bm
p}}\,\lambda^2\,\chi_{\phi\phi}({\bm k}+{\bm q},i\Omega), \nonumber\\
\label{eq:3.33d}
\eea
\ese

The development of the transport theory now proceeds analogously to the
electron-phonon scattering case, with
\be
\lambda^2\, \chi_{\phi\phi}({\bm p},i\Omega) \equiv V({\bm p},i\Omega)
\label{eq:3.34}
\ee
playing the role of the dynamical potential. The chief complications are, (1)
the spin structure, which leads to two independent matrix elements,
${\bm\Gamma}^{11}$ and ${\bm\Gamma}^{12}$ of the vertex function, and (2) the
fact that the vector vertex function ${\bm\Gamma}_{{\bm k}{\bm p}}$ depends on
the pitch vector ${\bm q}$ in addition to ${\bm k}$. (${\bm p}$ gets eliminated
by the Kronecker-$\delta$ constraints in Eqs.\ (\ref{eq:3.17a}) and
(\ref{eqs:3.33}).) In addition, there are two Fermi surfaces, a feature the
helimagnetic problem shares with the ferromagnetic one.

As a consequence of the dependence of the vertex function on ${\bm q}$, the
polarization function, and hence the conductivity, is no longer isotropic. The
tensor is still diagonal, but the transverse and longitudinal components, with
respect to ${\bm q}$, are different,
\be
\pi_{ij}(i\Omega) = \delta_{ij}\left[\left(\delta_{i1} +
\delta_{i2}\right)\pi_{\text{T}}(i\Omega) +
\delta_{i3}\pi_{\text{L}}(i\Omega)\right]
\label{eq:3.35}
\ee
As a consequence of the two Fermi surfaces, the conductivity is the sum of two
contributions,
\be
\sigma_{\text{L},\text{T}} = \Re \lim_{\omega\to
0}\sigma_{\text{L},\text{T}}(i\Omega \to \omega + i0) =
\sigma_{\text{L},\text{T}}^{(1)} + \sigma_{\text{L},\text{T}}^{(2)}.
\label{eq:3.36}
\ee
After calculations that are quite involved, but follow the same reasoning as in
the electron-phonon case, one obtains the following expression for the
transverse conductivity,
\bse
\label{eqs:3.37}
\bea
\sigma_{\text{T}}^{(i)} &=& \frac{e^2}{2m_{\text{e}}^2} \int
d\epsilon\,\left(\frac{-\partial
f}{\partial\epsilon}\right)\,\frac{1}{V}\sum_{\bm k} \delta\left(\omega_i({\bm
k})\right)\,k_{\perp}^2\,
\nonumber\\
&&\hskip 50pt \times\tau_i({\bm k},\epsilon)\,\Lambda_i({\bm k},\epsilon).
\label{eq:3.37a}
\eea
Here $\tau_i$ is the quasi-particle relaxation rate given by Eqs.\
(\ref{eqs:3.22}), $\omega_i$ are the resonance frequencies given by Eq.\
(\ref{eq:2.8}), and $f(\epsilon,T) = 1/\exp(\epsilon/T+1)$ is the Fermi
distribution function. The scalar vertex functions $\Lambda_i$ are related to
the part of the vector vertex function ${\bm\Gamma}_{{\bm k}{\bm p}}$ that is
proportional to ${\bm k}$. As in the electron-phonon case, they obey integral
equations which reduce to algebraic equations for the purpose of determining
the leading temperature dependence only of the conductivity. We obtain
\begin{widetext}
\be
\Lambda_i({\bm k},\epsilon) = 1 - \Lambda_i({\bm
k},\epsilon)\,\frac{2\tau_i({\bm k},\epsilon)}{\left(\xi_{{\bm k}+{\bm q}} +
\xi_{\bm k}\right)^2} \int du\,\left[n(u) + f(\epsilon +
u)\right]\,\frac{1}{V}\sum_{\bm p} g_i({\bm k},{\bm p}\,;\epsilon)\,f({\bm
k},{\bm p})\, V''({\bm k}-{\bm p},u)\,\delta(u - \omega_i({\bm p})),
\label{eq:3.37b}
\ee
\end{widetext}
where
\be
f({\bm k},{\bm p}) = 1 - \frac{1}{2k_{\perp}^2}\,({\bm k}-{\bm p})_{\perp}^2.
\label{eq:3.37c}
\ee
\ese
Substituting the solution of Eq.\ (\ref{eq:3.37b}) into Eq.\ (\ref{eq:3.37a})
shows that the conductivity is given in terms of transport relaxation time
$\tau_{\text{tr}}$ that is different from the quasi-particle relaxation time
$\tau$. Namely,
\bse
\label{eqs:3.38}
\be
\sigma_{\text{T}}^{(i)} = \frac{e^2}{2m_{\text{e}}^2} \int
d\epsilon\,\left(\frac{-\partial
f}{\partial\epsilon}\right)\,\frac{1}{V}\sum_{\bm k} \delta\left(\omega_i({\bm
k})\right)\,k_{\perp}^2\,\tau_{\text{tr}}^{(i)}({\bm k},\epsilon),
\label{eq:3.38a}
\ee
with
\bea
\frac{1}{2\tau_{\text{tr}}^{(i)}({\bm k},\epsilon)} &=&
\frac{-2\pi}{\left(\xi_{{\bm k}+{\bm q}} + \xi_{\bm k}\right)^2} \int
\frac{du}{\pi}\,\left[n(u) + f(\epsilon + u)\right]\,
\nonumber\\
&&\hskip -20pt \times\frac{1}{V}\sum_{\bm p} g_i({\bm k},{\bm p})\, V_2''({\bm
k}-{\bm p},u)\,\delta(u - \omega_i({\bm p})).
\nonumber\\
\label{eq:3.38b}
\eea
Here
\be
V_2''({\bm k}-{\bm p},u) = \frac{({\bm k}-{\bm
p})_{\perp}^2}{2k_{\perp}^2}\,V''({\bm k}-{\bm p},u).
\label{eq:3.38c}
\ee
\ese
From Eqs.\ (\ref{eqs:3.33}) and (\ref{eq:3.21c}) it follows that $({\bm k}-{\bm
p})_{\perp}^2 \sim u \sim T$ in a scaling sense. The leading temperature
dependence of the transport relaxation rate $1/\tau_{\text{tr}}$ will therefore
carry one higher power of the temperature than the quasi-particle relaxation
rate $1/\tau$.

The leading contribution to the longitudinal conductivity is given by Eq.\
(\ref{eq:3.37a}), with $k_{\perp}^2$ replaced by $k_z^2$. Since ${\bm k}$ is
pinned to the Fermi surface by the $\delta$-function in Eq.\ (\ref{eq:3.37a}),
this difference does not change the temperature dependence. The leading
temperature dependence of $\sigma_{\text{L}}$ is thus also again given by
$\tau_{\text{tr}}$.

Combining these results and observations with the results of Sec.\
\ref{subsubsec:III.B.3} we find the following asymptotic temperature dependence
for both the transverse and longitudinal transport relaxation rates due to
helimagnons in the ordered phase of a helimagnet with cubic symmetry:

\bse
\label{eqs:3.39}
\be
1/\tau_{\text{tr}}^{\text{T}} \propto 1/\tau_{\text{tr}}^{\text{L}} \propto
\left(\frac{q}{k_{\text{F}}}\right)^8
\left(\frac{\epsilon_{\text{F}}}{\lambda}\right)^2\left[ \nu^4
\left(\frac{T}{T_q}\right)^{5/2} \hskip -8pt +
\left(\frac{T}{T_q}\right)^{7/2}\right].
\label{eq:3.39a}
\ee
Here $\nu$ is the prefactor of the cubic anisotropy in Eq.\ (\ref{eq:3.25}),
$T_q$ is defined after Eq.\ (\ref{eq:3.8}), and we have again omitted all
numerical prefactors. For both the longitudinal and transverse resistivities,
this leads to Eq.\ (\ref{eq:2.15}) with
\be
\rho_{5/2} \propto \nu^4\,\lambda \left(\frac{q}{k_{\text{F}}}\right)^8\,
\left(\frac{\epsilon_{\text{F}}}{\lambda}\right)^2\, \frac{1}{T_q^{5/2}}\ .
\label{eq:3.39b}
\ee
\ese


\section{Discussion and Conclusion}
\label{sec:IV}

In summary, we have calculated the effects of the Goldstone mode in the ordered
phase of helical magnets, the helimagnon, on the low-temperature behavior of
the specific heat, the quasi-particle relaxation rate, and the resistivity. The
detailed microscopic calculations given in Sec.\ \ref{sec:III} have
corroborated the simple physical plausibility arguments given in Sec.\
\ref{sec:II}. The helimagnon contribution to the specific heat was found to
have a $T^2$ temperature dependence, whereas the corresponding contribution
from ferromagnetic and antiferromagnetic Goldstone modes goes as $T^{3/2}$ and
$T^3$, respectively.\cite{Kittel_1996} The quasi-particle relaxation rate was
found to have a $T^{3/2}$ temperature dependence, which is nonanalytic and
stronger than the $T^2$ behavior in a Fermi liquid. The resistivity depends on
the transport relaxation rate, whose temperature dependence is one power weaker
than that of the quasi-particle rate. The asymptotic low-temperature
contribution to the resistivity due to electron-helimagnon scattering thus goes
as $T^{5/2}$.

We divide our discussion of these results into a semi-quantitative discussion
that makes predictions of experimental relevance, and more general theoretical
remarks.

\subsection{Predictions for Experiments}
\label{subsec:IV.A}

The leading low-temperature results given in Sec.\ \ref{sec:III} hold for
temperatures $T\ll T_q$, with $T_q$ the temperature scale given after Eq.\
(\ref{eq:3.8}). This can be seen as follows. The helimagnon dispersion relation
as given in Eq.\ (\ref{eq:2.1}), or (\ref{eq:3.6'b}), is valid for wave numbers
$k<q$, and crosses over to a different behavior around $k=q$. The energy or
temperature scale related to this crossover is thus given by
\bse
\label{eqs:4.1}
\be
T_q \equiv \omega_0(k_z=q,k_{\perp}=0) = \sqrt{c_2}\,
\lambda\,q^2/k_{\text{F}}^2.
\label{eq:4.1a}
\ee
Here we have defined a dimensionless number $c_2$ by $c_z = c_2\,\lambda^2
q^2/k_{\text{F}}^4$. Within our weak-coupling model calculation we have $c_2 =
1/36$, or
\be
T_q = \lambda\,q^2/6 k_{\text{F}}^2,
\label{eq:4.1b}
\ee
\ese
which is also the energy of a ferromagnon at $k=q$ within Stoner theory.
Estimates for the parameters entering $T_q$ that are appropriate for MnSi have
been given in I: $q/k_{\text{F}} \approx 0.024$, and $\epsilon_{\text{F}}
\approx 23,000\,{\text{K}}$. The value of $\lambda$ is less certain, for
reasons explained in I. An upper limit is provided by the Fermi energy, and we
use $\lambda/\epsilon_{\text{F}} = 1/2$ as a practical upper bound. A likely
lower limit is given by $\lambda =
520\,{\text{K}}$.\cite{Taillefer_Lonzarich_Strange_1986} Given the large
ordered moment in MnSi,\cite{Pfleiderer_Julian_Lonzarich_2001} a value of
$\lambda$ that is a sizeable fraction of $\epsilon_{\text{F}}$ is actually more
plausible than the latter. For these two extreme choices one obtains $T_q
\approx 1.1\,{\text{K}}$ and $T_q \approx 50\,{\text{mK}}$, respectively. In
providing such estimates one should keep in mind that even semi-quantitative
estimates are difficult for a system like MnSi that is characterized by a
complex Fermi surface and strong interactions.\cite{correlations_footnote} It
should also be noted that $T_q$ gives only the general scale for the crossover;
a numerical evaluation of, for instance, the integral that determines the
specific heat, Eq.\ (\ref{eq:3.6}), shows clear deviations from the $T^2$
behavior already at a temperature of about $0.2 T_q$. The relevant temperature
scale for the interesting helimagnon effects is thus quite low.

To see what the behavior crosses over to at $T\approx T_q$, we realize that the
term $k_{\perp}^4$ in Eq.\ (\ref{eq:3.6'b}) is the leading (in a scaling sense)
contribution to a term $k^4$. For wave numbers $k$ larger than $q$, the
dispersion relation is given by $\omega_0({\bm k}) \propto k^2$, and $k_z$ and
$k_{\perp}$ both scale effectively as $T^{1/2}$. For the specific heat at
$T>T_q$ this implies a helimagnon contribution proportional to $T^{3/2}$. This
is the same behavior as in a ferromagnet, since the dispersion relation is
effectively ferromagnet-like and the specific heat depends only on the
dispersion relation.

A scaling expression for the helimagnon contribution to the specific heat that
incorporates both sides of this crossover is
\be
C_V^{\text{hm}}(T) = q^3\,f_C(T/T_q).
\label{eq:4.1'}
\ee
Here $f_C$ is a universal scaling function with $f_C(x\to 0) = A_C\,x^2$ and
$f_C(x\gg 1) = B_C\,x^{3/2}$. Here $A_C$ is given after Eq.\ (\ref{eq:3.8}),
and $B_C$ is another universal number.

It is also interesting to compare the helimagnon contribution to the
Fermi-liquid one. In the case of the specific heat, this means comparing the
first term in Eq.\ (\ref{eq:2.3}) with the second one. In a nearly-free
electron model, the coefficient $\gamma$ of the Fermi-liquid term in Eq.\
(\ref{eq:2.3}) is given by $\gamma =
k_{\text{F}}^3/6\epsilon_{\text{F}}$.\cite{Landau_Lifshitz_V_1980} From Eq.\
(\ref{eq:3.8}), we have $\gamma_2 = q^3 A_C/T_q^2$. The helimagnon contribution
is equal to the Fermi-liquid one at a temperature
\be
T^* = \gamma/\gamma_2 = \frac{1}{36 A_C}\,\frac{\lambda}{\epsilon_{\text{F}}}\,
\frac{k_{\text{F}}}{q}\, T_q.
\label{eq:4.2}
\ee
With the value of $k_{\text{F}}/q$ appropriate for MnSi given above, and
$\lambda$ a sizeable fraction of $\epsilon_{\text{F}}$, one finds $T^* \approx
T_q$; for smaller values of $\lambda$, $T^*$ is correspondingly smaller than
$T_q$ (which itself has a smaller value). Notice that $T<T^*$ does not preclude
an observation of the $T^2$ helimagnon contribution to the specific heat; it
can be extracted by plotting $C/T$ versus $T$.

For the quasi-particle relaxation rate, which can be measured by means of
either weak-localization or tunnelling experiments, a similar discussion
applies. The results given in Sec.\ \ref{subsubsec:III.B.3} apply for
temperatures small compared to $T_q$, and at higher temperatures a crossover
occurs to a regime that is characterized by an effectively quadratic helimagnon
dispersion relation and $k_z \sim k_{\perp} \sim T^{1/2}$. Repeating the
analysis in Sec.\ \ref{subsubsec:III.B.3} shows that in this regime,
$1/\tau({\bm k}) \propto T$. (We note in passing that this is {\em not} the
correct result for a ferromagnet, where the Goldstone dispersion relation is
effectively the same, but the structure of the equation analogous to Eq.\
(\ref{eq:3.21a}) is different.) This is true irrespective of whether one uses
the nearly-free electron energy-momentum relation of Sec.\ \ref{par:III.B.3.a}
of the cubic one of Sec.\ \ref{par:III.B.3.b}. This linear temperature
dependence is remarkable, although this is not the asymptotic low-temperature
contribution, since it deviates so strongly from the quadratic Fermi-liquid
result.

A scaling expression that incorporates this limit for both helimagnon
contributions to the relaxation rate is
\be
\frac{1}{\tau (\bm{k})} = \lambda\,\left(\frac{q}{k_{\text{F}}}\right)^6
\left(\frac{\epsilon_{\text{F}}}{\lambda}\right)^2 \left(
f_{\tau}^{(5/2)}(T/T_q) + f_{\tau}^{(3/2)}(T/T_q)\right),
\label{eq:4.2'}
\ee
with $f_{\tau}^{(5/2)}(x\to 0) \propto x^{5/2}$, $f_{\tau}^{(3/2)}(x\to 0)
\propto x^{3/2}$, and $f_{\tau}^{(5/2)}(x\gg 1) \propto f_{\tau}^{(3/2)}(x\gg
1) \propto x$.

Again, it is instructive to compare with the corresponding Fermi-liquid result,
which is
\be
\frac{1}{\tau_{\text{FL}}} =
\frac{\pi^3}{8}\,\epsilon_{\text{F}}\left(\frac{T}{\epsilon_{\text{F}}}\right)^2.
\label{eq:4.3}
\ee
We start with the $T^{5/2}$ term, Eq.\ (\ref{eq:3.24b}), which will be present
even in systems with a sizeable lattice anisotropy. It is larger than the
Fermi-liquid contribution for temperatures larger than a crossover temperature
\be
T^*_{2-5/2} = \frac{3\pi^6}{2A_{\tau}^2}\,
\left(\frac{k_{\text{F}}}{q}\right)^4
\left(\frac{\lambda}{\epsilon_{\text{F}}}\right)^6 T_q.
\label{eq:4.4}
\ee
Due to the high power of the uncertain parameter $\lambda/\epsilon_{\text{F}}$
that enters Eq.\ (\ref{eq:4.4}) it is hard to give even a semi-quantitative
estimate. For $\lambda$ a sizeable fraction of $\epsilon_{\text{F}}$ this
result suggests that $T^*_{2-5/2}$ is substantially larger than $T_q$, which
means that the $T^{5/2}$ contribution, in the region of its validity, is small
compared to the Fermi-liquid term. However, for system with a smaller value of
$\lambda/\epsilon_{\text{F}}$ the $T^{5/2}$ contribution may be important. We
now turn to the $T^{3/2}$ law that is the asymptotic helimagnon contribution to
the relaxation rate for a cubic system. By comparing Eqs.\ (\ref{eqs:3.28})
with Eq.\ (\ref{eq:4.3}) one finds that the $T^{3/2}$ law becomes comparable to
the Fermi-liquid $T^2$ behavior at a temperature
\be
T^*_{3/2 - 2} = 6\left(\frac{8B\nu^4}{\pi^3}\right)^2
\left(\frac{q}{k_{\text{F}}}\right)^4
\left(\frac{\epsilon_{\text{F}}}{\lambda}\right)^6 T_q
\label{eq:4.5}
\ee
with $B$ from Eq.\ (\ref{eq:3.28c}). The $T^{3/2}$ term is numerically large
compared to the Fermi liquid $T^2$ contribution for temperatures
$T<T^*_{3/2-2}$. The numerical value of $T^*_{3/2 - 2}$ depends both on a high
power of the parameter $\nu$ in addition to the high power of
$\epsilon_{\text{F}}/\lambda$. With $\nu=1$ and $\epsilon_{\text{F}}/\lambda=2$
as above one finds $T^*_{3/2-2} \approx 0.025\, T_q$. Smaller values of
$\lambda$ result in a larger ratio $T^*_{3/2-2}/T_q$. $T^*_{3/2-2} \agt T_q$ is
necessary for the $T^{3/2}$ term to be numerically large compared to the
Fermi-liquid contribution in the entire range of its validity, and for the
change from the asymptotic $T^{3/2}$ behavior to the pre-asymptotic $T$ term to
occur at a temperature low enough for the helimagnon contribution to be still
large compared to the Fermi-liquid contribution. According to the above
estimate this is the case for $\lambda \alt \epsilon_{\text{F}}/4$.

At this point we need to remember that the helimagnon dispersion relation we
have used to calculate the relaxation rate, Eq.\ (\ref{eq:2.1}), is valid only
for rotationally invariant systems, while the $T^{3/2}$ behavior is the
consequence of an anisotropic electron dispersion relation, Eq.\
(\ref{eq:3.25}). However, since the modification of the helimagnon dispersion
is small, namely, on the order of the spin-orbit coupling squared, there still
is a large temperature range where the $T^{3/2}$ behavior is realized. To see
this, we recall that the spin-orbit coupling $g_{\text{so}}$, which is
proportional to $q$, changes Eq.\ (\ref{eq:2.1}) to (see I Eq. (2.23))
\be
\omega_0({\bm k}) = \sqrt{c_z k_z^2 + c_{\perp} k_{\perp}^4 + b\,c_z q^2
k_{\perp}^2/k_{\text{F}}^2}\ ,
\label{eq:4.6}
\ee
where $b$ is a number on the order of unity. If we scale $k_z$ and $k_{\perp}$
with appropriate powers of $T$, $c_z$, and $c_{\perp}$, such that the main
contribution to the integral in Eq.\ (\ref{eq:3.21a}) comes from $k_z \approx
k_{\perp} \approx 1$, we see that the spin-orbit term in $\omega_0$ is
unimportant for temperatures
\be
T > T_{\text{so}} = b\,\lambda (q/k_{\text{F}})^4 \propto (q/k_{\text{F}})^2
T_q.
\label{eq:4.7}
\ee
On the other hand, the $T^{3/2}$ term dominates over the $T^{5/2}$ term for
temperatures
\be
T < T^*_{3/2-5/2} = \sqrt{T^*_{3/2-2} T^*_{2-5/2}} =
\frac{8B\nu^4}{A_{\tau}}\left(\frac{q}{k_{\text{F}}}\right)^2\lambda.
\label{eq:4.8}
\ee
As expected, $T_{\text{so}}$ is small compared to $T^*_{3/2-5/2}$ by a factor
of $(q/k_{\text{F}})^2 \propto g_{\text{so}}^2$.

Finally, the helimagnon contribution to the resistivity is suppressed compared
to the contribution to the quasi-particle scattering rate by a factor of
$T/\lambda$. For realistic temperatures one thus expects the Fermi-liquid $T^2$
contribution to dominate. This is consistent with the statement in Ref.\
\onlinecite{Pfleiderer_Julian_Lonzarich_2001} that in the ordered phase of
MnSi, the resistivity shows a $T^2$ behavior.

\subsection{General remarks}
\label{subsec:IV.B}

All of our results hinge on the Goldstone mode not being overdamped. In an
ultraclean system, a generic band structure actually leads to a weak
overdamping of the mode in certain directions in momentum space, which can
change the above power laws at very low temperatures.\cite{AR_unpublished}
However, even small amounts of quenched disorder qualitatively {\em weaken} the
damping, see Sec. IV.E.2 and Ref. 34 in I. The results also hinge on rotational
invariance. If the direction of the helix is pinned by the underlying lattice,
then the $\sqrt{k_z^2 + k_{\perp}^4}$ dispersion relation crosses over to a
linear one at asymptotically small wave numbers, and the temperature
dependences of the observables will cross over to the same power laws as in the
acoustic phonon case. The crossover temperature for this effect to become
relevant depends on the strength of the pinning.

For the specific heat we have found that the asymptotic low-temperature
dependence is in between the one for ferromagnets and antiferromagnets,
respectively, as one might expect from the fact that the helimagnon dispersion
relation, Eq.\ (\ref{eq:2.1}), is antiferromagnet-like in the longitudinal
direction, and ferromagnet-like in the transverse one. We point out, however,
that the Goldstone susceptibility at zero frequency diverges more strongly with
vanishing wave vector ${\bm k}$ than in the ferromagnetic case. Namely, from
Eq.\ (\ref{eq:3.22b}) (or from I Eq. (4.33a)) we see that in the helimagnetic
case the static susceptibility diverges as $1/(k_z^2 + {\bm k}_{\perp}^4)$,
whereas in a ferromagnet the corresponding behavior is $1/{\bm k}^2$. One
expects this to have dramatic effects for the hydrodynamics, as is the case in
certain liquid crystals.\cite{Mazenko_Ramaswamy_Toner_1983} This problem will
be studied separately.

The transport theory presented in this paper is still at a rather primitive
level. We have calculated the resistivity in the simplest possible consistent
approximation, which corresponds to a solution of a Boltzmann equation where it
is assumed that the bosonic modes remain in thermal equilibrium. Whether this
zero-loop approximation yields indeed the leading low-temperature behavior of
the resistivity is an open question. In ferromagnets, mode-mode coupling
effects (or loops in a field-theoretic language) have been found to induce a
square-root frequency dependence in the zero-temperature
resistivity.\cite{Kirkpatrick_Belitz_2000} Possible similar effects in the
helimagnetic case, and the nature of the mode-mode coupling effects at nonzero
temperature, remain to be explored. The results of the current calculation,
namely, effects of the helimagnons on the conductivity that are weaker than the
usual $T^2$ Fermi-liquid behavior, are consistent with reports of a $T^2$
behavior of the resistivity in the ordered phase of
MnSi.\cite{Pfleiderer_Julian_Lonzarich_2001} However, transport in the ordered
phase has not been investigated systematically, and the observation of possible
mode-mode coupling effects may require measurements at very low temperatures.

The quasi-particle relaxation rate $1/\tau$ has the somewhat surprising
property that its temperature dependence depends on the details of the
electronic energy-momentum relation $\epsilon_{\bm k}$. As was shown in Sec.\
\ref{subsubsec:III.B.3}, the generic behavior for an $\epsilon_{\bm k}$
consistent with a cubic symmetry is $1/\tau \propto T^{3/2}$, whereas a
nearly-free electron model leads to $\tau \propto T^{5/2}$ for generic ${\bm
k}$. This dependence of asymptotic low-temperature properties on microscopic
details is unusual and raises questions about universality. We also note that
the prefactor of the leading $T^{3/2}$ behavior depends on the fourth power of
the parameter $\nu$ that describes deviations from a nearly-free electron
model, see Eq.\ (\ref{eq:3.28c}). In systems where the deviations from
nearly-free electron behavior are small, the asymptotic $T^{3/2}$ law could
then be confined to extremely low temperatures.

We stress that the current theory deals with the {\em ordered} phase of a
helimagnet, whereas experimentally much more interesting behavior, namely, a
$T^{3/2}$ behavior of the resistivity in violation of Fermi-liquid theory, has
been observed in the {\em disordered}
phase.\cite{Pfleiderer_Julian_Lonzarich_2001} While the current theory has
nothing to say about this, the observed remnants of helical order in at least
part of the region where the non-Fermi-liquid behavior is
observed\cite{Pfleiderer_et_al_2004} makes an understanding of the ordered
phase a likely necessary prerequisite for a theoretical treatment of the
disordered phase.

Finally, we mention that the results for the quasi-particle relaxation rate and
the resistivity presented above hold for clean systems. The presence of
quenched disorder can drastically change the transport behavior; this will be
investigated in a separate publication.\cite{us_tbp}

\acknowledgments We thank the participants of the Workshop on Quantum Phase
Transitions at the KITP at UCSB, and in particular Christian Pfleiderer and
Thomas Vojta, for stimulating discussions. This work was supported by the NSF
under grant Nos. DMR-01-32555, DMR-01-32726, PHY99-07949, DMR-05-29966, and
DMR-05-30314, and by the DFG under SFB 608.

\appendix

\section{Matsubara frequency sums}
\label{app:A}

In this appendix we derive the identity (\ref{eq:3.4}). Consider an even
function $f(z) = f(-z)$ of one complex argument $z$. Let $f$ have singularities
(poles or cuts) on the real axis, but be analytic for $\Im z > 0$. Let
$\Omega_n = 2\pi Tn$ ($n$ integer) be the set of bosonic Matsubara frequencies.
Using standard techniques the functional
\bse
\label{eqs:A.1}
\be
F = T\sum_{n=1}^{\infty} f(i\Omega_n)
\label{eq:A.1a}
\ee
can be written
\be
F = \frac{-i}{2\pi} \int_{-\infty}^{\infty} d\omega\,\nu(\omega)\,f(\omega +
i0),
\label{eq:A.1b}
\ee
where
\be
\nu(z) = \frac{1}{2}\,\coth \frac{z}{2T}\, - \frac{T}{z}\ .
\label{eq:A.1c}
\ee
\ese
The advantage of $\nu(z)$ over the more commonly used Bose distribution
function $n(z)$ is that $\nu(z)$ is odd in $z$ and has no pole at $z=0$. On the
real axis, $n$ has the property
\be
\nu(\omega) = T\,\frac{d}{d\omega}\,\left(\ln\sinh\frac{\vert\omega\vert}{2T} -
\ln\vert\omega\vert\right).
\label{eq:A.2}
\ee
Using this in Eq.\ (\ref{eq:A.1b}), and integrating by parts, yields
\be
F = \frac{-T}{\pi} \int_0^{\infty}
d\omega\,\left[\ln\sinh\frac{\vert\omega\vert}{2T} -
\ln\vert\omega\vert\right]\,\frac{\partial}{\partial\omega}\,\Im f(\omega +
i0).
\label{eq:A.3}
\ee
Only the imaginary part of $f(\omega + i0)$ contributes since the real part is
an even function of $\omega$. With $f = \ln\Gamma$ this yields Eq.\
(\ref{eq:3.4}).


\begin{thebibliography}{37}
\bibitem[{rot()}]{rot_invariance_footnote}
\bibinfo{note}{This holds for rotationally invariance systems. The spin-orbit
  coupling breaks this symmetry, since it couples the electron spins to the
  underlying lattice. We will discuss the consequences of this (weak) effect in
  Sec.\ \ref{subsec:IV.A}.}
\bibitem[{LRO()}]{LRO_footnote}
\bibinfo{note}{We note in passing that the true asymptotic low-temperature
  contribution to the specific heat is expected to be $T^2 \ln T$, since the
  strong fluctuations that result from the anisotropic helimagnon dispersion
  relation will not allow for true long-range order; see Ref. 16 in I. We
  ignore this effect here, as we did in I, as it will become relevant only at
  extremely long length scales and correspondingly low temperatures.}
\bibitem[{sca()}]{scaling_footnote}
\bibinfo{note}{In our notation, $a\propto b$ means ``$a$ is proportional to
  $b$'', and $a\sim b$ means ``$a$ scales like $b$''.}
\bibitem[{fm_()}]{fm_footnote}
\bibinfo{note}{For ferromagnets, the arguments given here do not apply, and we
  therefore do not discuss this case.}
\bibitem[{C_f()}]{C_footnote}
\bibinfo{note}{The usual specific heat $C_V$ is defined by this derivative with
  the volume and the number of particles kept fixed. Here we implicitly keep
  the volume and chemical potential fixed. For the leading term as
  $T\rightarrow 0$ these two operations give the same result.}
\bibitem[{dam()}]{damping_footnote}
\bibinfo{note}{This remains true if the wave vector dependence of the damping
  coefficient has a different functional form than the one in I Eq.\ (4.33c),
  as long as the mode remains propagating.}
\end{thebibliography}

\end{document}